\documentclass[12pt]{article}
\pdfoutput=1
\usepackage[a4paper]{geometry}
\usepackage{jheppub, amsmath,amssymb,amsfonts,amsxtra, mathrsfs, makeidx,graphics,graphicx,amsthm,epsfig, ytableau,bm,longtable,float, color,tikz,mathtools,xfrac,footnote,rotating, lscape, makecell, environ,mathtools, empheq, colortbl}

\usetikzlibrary{positioning}
\usetikzlibrary{chains}
\usetikzlibrary{arrows,fit,decorations.pathreplacing}
\tikzstyle{every picture}+=[remember picture]
\tikzstyle{na} = [baseline]

\usetikzlibrary{arrows, decorations.markings, calc, fadings, decorations.pathreplacing, patterns, decorations.pathmorphing, positioning}

\def\node#1#2{\overset{#1}{\underset{#2}{{\color{gray} \bullet}}}}

\def\sqwnode#1#2{\overset{#1}{\underset{#2}{{\square}}}}

\def\node#1#2{\overset{#1}{\underset{#2}{\circ}}}

\def\sqwnode#1#2{\overset{#1}{\underset{#2}{{ \square}}}}

\tikzstyle{every picture}+=[remember picture]
\tikzstyle{na} = [baseline=-.5ex]

\usepackage{bbm}
\usepackage{subfigure}

\newcommand{\eg}{\textit{e.g.}}

\newcommand{\ie}{\textit{i.e.}}

\numberwithin{equation}{section}

\newcommand{\be}{\begin{equation}} \newcommand{\ee}{\end{equation}}
\newcommand{\bea}{\begin{equation} \begin{aligned}} \newcommand{\eea}{\end{aligned} \end{equation}}

\def\tilde{\widetilde}

\def\hat{\widehat}

\def\bar{\overline}

\def\rt2{\sqrt{2}}

\def\mod{{\rm mod}}

\def\Tr{\mathop{\rm Tr}}
\def\tr{\mathop{\rm tr}}

\def\CI{{\cal I}}

\def\CN{{\cal N}}
\def\CO{{\cal O}}

\def\CT{{\cal T}}

\def\1{{\ds 1}}

\newcommand{\cH}{\mathcal{H}}
\newcommand{\cI}{\mathcal{I}}

\newcommand{\cN}{\mathcal{N}}

\def\repa{\raise4pt\hbox{$\square$}\mkern-14mu\raise-4pt\hbox{$\square$}}
\def\repab{\overline{\raise4pt\hbox{$\square$}\mkern-14mu\raise-4pt\hbox{$\square$}\mkern-1mu}}

\def\smileface{\ensuremath{\hbox{\large$\bigcirc$}\mkern-15mu\raise-1pt\hbox{\scriptsize$\smallsmile$}%
\mkern-10mu\raise4pt\hbox{..}\mkern4mu}}
\def\frownface{\ensuremath{\hbox{\large$\bigcirc$}\mkern-15mu\raise-1pt\hbox{\scriptsize$\smallfrown$}%
\mkern-10mu\raise4pt\hbox{..}\mkern4mu}}

\newcommand{\ba}{\begin{array}}
\newcommand{\ea}{\end{array}}
\newcommand{\bi}{\begin{itemize}}
\newcommand{\ei}{\end{itemize}}
\def\vec#1{\bm{#1}}
\def\bea#1\eea{\allowdisplaybreaks \begin{align}#1\end{align}}
 \newcommand{\ben}{\begin{enumerate}}
\newcommand{\een}{\end{enumerate}}
\newcommand{\bean}{\begin{eqnarray*}}
\newcommand{\eean}{\end{eqnarray*}}
\newcommand{\eref}[1]{(\ref{#1})}

\newcommand{\BC}{\mathbb{C}}
\newcommand{\BR}{\mathbb{R}}

\newcommand{\BZ}{\mathbb{Z}}

\newcommand{\BH}{\mathbb{H}}

\newcommand{\comment}[1]{}
\newcommand{\Sym}{\mathrm{Sym}}

\definecolor{light-gray}{gray}{0.7}

\newcommand{\red}{\color{red}}

\def\aup#1 {\overset{#1}{\uparrow} \, \overset{\tilde{#1}}{\downarrow}}

\tikzset{snake it/.style={decorate, decoration={snake, amplitude=.4mm, segment length=2mm,
                       post length=0mm,pre length=0mm}}}

\title{Supersymmetric Indices of 3d $S$-fold SCFTs}
\author[a,b]{Ivan Garozzo,} 
\author[a,b]{Gabriele Lo Monaco,} 
\author[b,c]{Noppadol Mekareeya,}
\author[a,b]{and \\ Matteo Sacchi}
\affiliation[a]{Dipartimento di Fisica, Universit\`a di Milano-Bicocca, \\ Piazza della indexenza 3, I-20126 Milano, Italy}
\affiliation[b]{INFN, sezione di Milano-Bicocca, \\Piazza della indexenza 3, I-20126 Milano, Italy}
\affiliation[c]{Department of Physics, Faculty of indexence, \\
Chulalongkorn University, Phayathai Road, \\
Pathumwan, Bangkok 10330, Thailand}
\emailAdd{ivangarozzo@gmail.com}
\emailAdd{gabriele.lomonaco92@gmail.com}
\emailAdd{n.mekareeya@gmail.com}
\emailAdd{m.sacchi13@campus.unimib.it}
\abstract{Enhancement of global symmetry and supersymmetry in the infrared is one of the most intriguing phenomena in quantum field theory.  We investigate such phenomena in a large class of three dimensional superconformal field theories, known as the $S$-fold SCFTs.  Supersymmetric indices are computed for a number of theories containing small rank gauge groups.  It is found that indices of several models exhibit enhancement of supersymmetry at the superconformal fixed point in the infrared. {Dualities} between $S$-fold theories that have different quiver descriptions are also analysed.  We explore a new class of theories with a discrete global symmetry, whose gauge symmetry in the quiver has a different global structure from those that have been studied earlier.}
\begin{document}
\maketitle

\section{Introduction}
Duality walls have played a central role in constructing a number of three dimensional superconformal field theories (SCFTs) with several rich properties \cite{Gaiotto:2008ak}.  Such objects can be introduced into a Type IIB brane configuration \cite{Hanany:1996ie} involving D3, NS5 and D5 branes, preserving eight supercharges.  The duality wall gives rises to a local $SL(2,\BZ)$ transformation at the position where it is located. The effective three dimensional description on the worldvolume of the D3 branes of such a system, with the inclusion of a duality wall, is interesting.  As pointed out in \cite{Gaiotto:2008ak, Gulotta:2011si}, the intersection between a stack of $N$ D3 branes and an $S$-duality wall gives rise to a 3d $\CN=4$ SCFT with a global symmetry $U(N) \times U(N)$, coupled to two $U(N)$ groups coming from the worldvolume of the D3 branes on each side of the wall.  The aforementioned SCFT was dubbed $T(U(N))$ in \cite{Gaiotto:2008ak} and we review some of its important features in section \ref{Sec:SusySfold}.  One may also consider the duality wall associated with a more general $SL(2,\BZ)$ element, for example $J=-ST^k$, where $S$ and $T$ are the generators of $SL(2,\BZ)$ such that $S^2=-1$ and $(ST)^3=1$.  The above effective description gets modified simply by turning on the Chern--Simons level $k$ to one of the $U(N)$ gauge groups.   Such an effective description is expected to flow to a non-trivial fixed point in the infrared, and the corresponding theory is referred to as an $S$-fold SCFT \cite{Assel:2018vtq}.  

The $S$-fold SCFT has an interesting holographic dual \cite{Assel:2018vtq}.  This involves $\mathrm{AdS}_4 \times M_6$ Type IIB string solutions with monodromies in $M_6$ in $SL(2,\BZ)$\footnote{The idea of using $SL(2,\BZ)$ monodromy to obtain new solutions was also applied in other dimensions. For example, those in $\mathrm{AdS}_5$ was considered in \cite{Garcia-Etxebarria:2015wns, Aharony:2016kai}, and those in $\mathrm{AdS}_3$ were considered in \cite{Couzens:2017way, Couzens:2017nnr}.}.  These solutions can be realised as the $SL(2,\BZ)$ quotient on the solutions which are the holographic dual of Janus interfaces in 4d $\CN=4$ super-Yang-Mills \cite{DHoker:2007zhm, DHoker:2007hhe}. The supergravity solution corresponding to the duality wall, dubbed the $S$-fold solution, was studied in \cite{Inverso:2016eet}.  We shall henceforth use the term duality wall associated with $J \in SL(2,\BZ)$ and $J$-fold interchangeably.   Several related realisations of duality walls in 4d $\CN=4$ super-Yang-Mills with $SL(2,\BZ)$ monodromies were also studied in \cite{Martucci:2014ema, Gadde:2014wma, Assel:2016wcr, Lawrie:2018jut}.  The abelian $S$-fold theories (without hypermultiplet matter) were investigated in \cite{Ganor:2014pha}.  {The moduli space of $S$-fold theories with unitary gauge groups was analysed in \cite{Garozzo:2018kra}, and those with gauge groups $SO(2N)$, $USp'(2N)$ and $G_2$ were analysed in \cite{Garozzo:2019hbf}.}

It is worth pointing out that the authors of \cite{Terashima:2011qi, Gang:2015wya, Gang:2018wek, Gang:2018huc} have studied (mainly in the context of the 3d--3d correspondence) very closed cousins of the $S$-fold theories \`a la \cite{Assel:2018vtq}. The difference between the two is that the gauge group is taken to be $SU(N)$ in the former, instead of $U(N)$ as in the latter. In this paper, we demonstrate that in the absence of hypermultiplet matter but with only $T(U(N))$ or $T(SU(N))$ components present in the effective description, such theories discussed in the former and in the latter are actually the same.

Another fascinating aspect of the $S$-fold SCFT is the amount of supersymmetry it possesses.  In the effective description of the $S$-fold theory, upon gauging the $U(N) \times U(N)$ symmetry of the $T(U(N))$ theory, supersymmetry is naively expected to be broken to $\CN=3$.  However, it was shown in \cite{Gang:2018huc, Assel:2018vtq} that in certain circumstances, supersymmetry can be enhanced to $\CN=4$ at the SCFT fixed point.

The main point of this paper is to study properties of $S$-fold SCFTs, including global symmetry and supersymmetry, using three dimensional supersymmetric index (or simply ``index'' for brevity) \cite{Bhattacharya:2008zy,Bhattacharya:2008bja, Kim:2009wb,Imamura:2011su, Kapustin:2011jm, Dimofte:2011py, Aharony:2013dha, Aharony:2013kma}. Such a quantity has proved extremely useful to study various intriguing aspects of quantum field theory, including enhancement of global symmetry and supersymmetry\footnote{The literature on the subject of supersymmetry enhancement is vast.  See, for example,~ \cite{Maruyoshi:2016tqk, Maruyoshi:2016aim, Agarwal:2016pjo, Benvenuti:2017lle, Benvenuti:2017kud, Agarwal:2017roi, Benvenuti:2017bpg, Agarwal:2018ejn, Gaiotto:2018yjh, Benini:2018bhk, Bashmakov:2018ghn, Giacomelli:2018ziv, Carta:2018qke, Gang:2018huc, Fazzi:2018rkr, Apruzzi:2018xkw, Agarwal:2018oxb, Aprile:2018oau} for some recent work in 3d and 4d. {See also \cite{Okazaki:2019ony} for a very recent paper that discussed supersymmetric indices of 3d $\CN=4$ gauge theories.}}.   Indeed, for 3d SCFTs, it is possible to put various short multiplets into equivalence classes according to how they contribute to the index \cite{Razamat:2016gzx} (see also \cite{Gadde:2009dj, Beem:2012yn}).  It also allows one to identify the current of the enhanced symmetry.  For theories with at least $\CN=3$ supersymmetry, including $S$-fold SCFTs, the index serves as a rather simple tool to diagnose the presence of the extra-supersymmetry current multiplet, which gives rise to the enhancement of supersymmetry (see \eg~ \cite{Evtikhiev:2017heo}).   Moreover, since the index is sensitive to the global structure of the gauge symmetry, it allows us to distinguish properties of $S$-fold theories with different gauge groups, such as $SU(N)$ versus $SU(N)/\BZ_N$.

The paper is organised as follows. In section \ref{Sec:SusySfold}, we provide a quick review of $S$-fold SCFTs.  In section \ref{Sec:Index}, three dimensional supersymmetric index is summarised in brief.  In subsection \ref{subsec:multiplets}, the contributions of various superconformal multiplets to the index are discussed.  In section \ref{sec:onegaugegroup}, we discuss $S$-fold theories with a single gauge group, both in the absence and in the presence of hypermultiplet matters.  We also study duality for a theory with two gauge groups and use index to understand the operator map between such theories.  In section \ref{sec:twoTtwogaugegroups}, we investigate theories corresponding to two duality walls and with two gauge groups.  The addition of fundamental hypermultiplet matter to such theories is discussed in subsection \ref{subsec:addflvtwoTtwonodes}.  
In section \ref{sec:SU2modZ2}, we consider theories with $SU(2)/\BZ_2$ gauge group with various Chern--Simons level and use the index to study the discrete $\BZ_2$ global symmetry of such theories.  Finally, we conclude the paper in section \ref{sec:conclusion} and discuss some open problems.

\section{3d $S$-fold SCFTs}
\label{Sec:SusySfold}
A large class of 3d $\mathcal{N}=4$ gauge theories admits a brane construction in terms of D3-branes, NS5-branes and D5-branes \cite{Hanany:1996ie}, spanning the following directions
\be
\begin{tabular}{c||cccccc|c|cccc}
\hline
~     & 0 & 1 &2 &3 & 4 & 5 & 6 & 7 &8 &9 &\\
\hline
D3   & X & X & X&  &  &   & X&  & & & \\ 
NS5 & X & X & X& X & X & X &    &   & & &\\
D5   & X & X & X&  &  &  &  &X &X &X & \\
\end{tabular}
\ee
In the following, we assume that the $x^6$ direction is compact.  Such a gauge theory admits a description in terms of the circular quiver with unitary gauge nodes, and possibly with fundamental hypermultiplets transforming under each gauge node.  A large number of such theories flow to interacting superconformal field theories (SCFTs).  The landscape of possible theories becomes even larger if we include new ingredients, such as duality walls, in the aforementioned brane configuration. 

As described by \cite{Gaiotto:2008ak, Gulotta:2011si, Assel:2014awa}, one can apply an action of $SL(2,\BZ)$, generated by $S$ and $T$ such that $S^2=-1$ and $(ST)^3=1$\footnote{We take the matrix representations of $S$ and $T$ to be $S= \begin{pmatrix} 0 & -1 \\ 1 & 0 \end{pmatrix}$ and $T= \begin{pmatrix} 1 & 0 \\ 1 & 1 \end{pmatrix}$.}, locally on the Type IIB brane system.  As an example, a $(p, q)$ fivebrane is transformed into a $(-q, p)$ fivebrane under $S$-transformation.  In terms of duality walls, such a local $S$-transformation can be formulated as follows: a $(p, q)$ fivebrane can be traded for a $(-q, p)$ fivebrane with an $S$-duality wall on its right and $(S^{-1})$-duality wall on its left.  In this sense, the duality walls do not only act as the boundaries of the region of the local $SL(2,\BZ)$ action, but each of them can also be regarded as a new object in the Type IIB brane configuration.  In this paper, we refer to the 3d effective theory associated with such a brane system, along with the duality walls, as an {\it $S$-fold theory}.  Such a theory is believed to flow to a conformal field theory, which we refer to as {\it $S$-fold SCFT}, in the infrared.

Let us first explore the system containing only D3 branes and a duality wall associated with an $SL(2,\BZ)$ element $J$ (also known as the $J$-fold), in the absence of NS5 and D5 branes. The duality wall gives rise to the $J$-twisted boundary conditions in the 4d $\CN=4$ super-Yangs-Mills (SYM), which is the worldvolume of the D3 branes, on the circle.  Tuning the holomorphic coupling $\tau$ of the SYM theory such that it varies in a small neighbourhood of the duality wall and almost constant elsewhere, one sees that the ultraviolet description of such a configuration is 4d $\CN=4$ SYM on a circle coupled to a certain 3d theory associated with the monodromy $J$. As pointed out in \cite{Gaiotto:2008ak}, for $J=-ST^k$, the latter is identified with a 3d $\CN=4$ SCFT known as the $T(U(N))$ theory, whose flavour symmetry is $U(N) \times U(N)$, with one of the $U(N)$ groups having Chern--Simons (CS) level\footnote{Throughout the paper, $G_k$ denotes gauge group $G$ with Chern-Simons level $k$.} $k$.  The 3d effective description of $N$ D3-branes with a duality wall associated with $J=-ST^k$ can therefore be written in terms of the following quiver diagram:
\be \label{quiveroneloop}
\begin{tikzpicture}[baseline]
\tikzstyle{every node}=[font=\footnotesize, node distance=0.45cm]
\tikzset{decoration={snake,amplitude=.4mm,segment length=2mm,
                       post length=0mm,pre length=0mm}}
\draw[blue,thick] (0,0) circle (1.5cm) node[midway, right] {$N$ D3};
\draw[decorate,red,thick] (0,1) -- (0,2) node[right] {$J_k= - ST^k$};
\end{tikzpicture}
\qquad \qquad \qquad
\begin{tikzpicture}[baseline]
\tikzstyle{every node}=[font=\footnotesize]
\node[draw, circle] (node1) at (0,0) {$N_k$};
\draw[red,thick] (node1) edge [out=45,in=135,loop,looseness=5, snake it]  (node1);
\node[draw=none] at (1.4,0.5) {{\red $T(U(N))$}};
\end{tikzpicture}
\ee
In the brane picture (left figure), we denote the duality wall by the wiggly line.  In the quiver diagram (right figure), the circular node with label $N_k$ denotes the $U(N)_k$ gauge group, and the red wiggly line denotes the $T(U(N))$ theory.  (Throughout the paper, we shall refer to the red wiggly line in the quiver diagram as a $T$-link.)  Another equivalent way to represent the above quiver diagram is to write $T(U(2))/U(2)^{\text{diag}}_{k}$, where the diagonal subgroup $U(2)^\text{diag}$ of the symmetry $U(2) \times U(2)$ of $T(U(2))$ is gauged with CS level $k$.

Let us briefly review some properties of the $T(U(N))$ theory \cite{Gaiotto:2008ak}.  This theory can be constructed from the $T(SU(N))$ theory, which  can be described in terms of the folllowing 3d $\CN=4$ quiver gauge theory:
\be \label{12N} 
\node{}{1}-\node{}{2}- \cdots - \node{}{N-1}-\sqwnode{}{N}
\ee
where each circular node labelled by $m$ denotes the gauge group $U(m)$.  The flavour symmetry algebra of this theory is $su(N)$, as manifest in the quiver diagram; this is the symmetry of the Higgs branch of \eref{12N} .  On the other hand, the $U(1)^{N-1}$ topological symmetry of this theory gets enhanced to $su(N)$ in the infrared; this is the symmetry of the Coulomb branch of \eref{12N}. In order to obtain the $T(U(N))$ theory, one introduces the so-called $T(U(1))$ theory \cite{Gaiotto:2008ak}, which is an almost empty theory that contains the $U(1) \times U(1)$ background vector multiplets with the mixed CS level $-1$.  The product of $T(U(1))$ and $T(SU(N))$ is then the $T(U(N))$ theory.  Since only one of the $su(N)$ symmetry of \eref{12N} is manifest in the Lagrangian description, quiver \eref{quiveroneloop} with $N>1$ should be regarded as a ``quasi-Langrangian'' description of the 3d theory in question.

Another interesting point regarding quiver \eref{quiveroneloop} is the amount of supersymmetry.  Since the diagonal subgroup of the Higgs and Coulomb branch symmetries of $T(U(N))$ is gauged, the $SU(2) \times SU(2)$ $R$-symmetry of the $\CN=4$ theory is naively expected to be broken to a diagonal subgroup $SU(2)$, which corresponds to $\CN=3$ supersymmetry.  However, as pointed out in \cite{Gang:2018huc} (using the $SU(2)$ gauge group and $|k|=3$) and later in \cite{Assel:2018vtq} (particularly for the large $N$ limit and $|k|>2$), supersymmetry can be enhanced to $\CN=4$ in certain cases.

One way to generalise \eref{quiveroneloop} is to introduce more than one duality wall, say those associated with $J_{k_1} = -ST^{k_1}, \, J_{k_2}=-ST^{k_2}, \, \ldots, \, J_{k_n}=-ST^{k_n}$ into the system consisting of $N$ D3-branes.  (An example for $n=2$ and $N=2$ is depicted in \eref{twoTlinksnoflv}.)  This system of duality walls corresponds to the $SL(2,\BZ)$ element $J = \prod_{i=1}^n J_{k_i}$, and this can be classified according to the value of $|\Tr J|$.  If $|\Tr J| <2$, then $J$ is said to be elliptic. If $|\Tr J| >2$, then $J$ is said to be parabolic.  If $|\Tr J|=2$, then $J$ is said to be hyperbolic.  The abelian case of $N=1$, with $J$ hyperbolic, has been extensively studied in \cite{Ganor:2014pha}.   More generally, the authors of \cite{Assel:2018vtq} studied the holographic dual of such a system as well as the three sphere partition function in the large $N$ limit and with $J$ hyperbolic, and showed that supersymmetry of the theory is enhanced to $\CN=4$.  It should be noted that any parabolic element of $SL(2,\BZ)$ is conjugate to $\pm T^p$, for some $p \neq 0$; since this does not involve $S$, it is expected that the corresponding theory is simply 3d $\CN=3$ CS theory with the gauge group $U(N)$\footnote{We find that the supersymmetric indices for the hyperbolic cases diverge. However, when we add hypermultiplet matter to such theories, the indices are well-defined and the theory have some interesting physical properties, as we shall discuss below.}, which upon integrating out the auxiliary fields, one obtains a pure CS theory.

An interesting further generalisation of the above system is to add NS5 branes and D5 branes\footnote{In \cite{Assel:2018vtq}, the authors also studied theories that arise from such a brane system, with an $S$-duality wall.  According to their finding, this boundary condition has to be accompanied by a flip of the coordinates $(x^{3,4,5}\,,\,x^{7,8,9})\,\rightarrow\,(x^{7,8,9}\,,\,-x^{3,4,5})$.  As a result, such a configuration is referred to as an $S$-flip. It was conjectured in \cite{Assel:2018vtq} that the theory with an $S$-flip (along with bifundamental and fundamental hypermultiplets) has $\CN=3$ supersymmetry (at least in the large $N$ limit).}.  In terms of the quiver description, this corresponds to adding bifundamental and fundamental hypermultiplets to theories with $T$-links.  In this paper, we examine the system at finite $N$ (in particular, $N=1$ and $N=2$) and find that the supersymmetric indices of such theories exhibit supersymmetry enhancement for certain CS levels.  We also observe that upon adding hypermultiplet matter to hyperbolic $J$-fold theories, the resulting theory becomes highly non-trivial, mainly due to the presence of the gauge neutral monopole or dressed monopole operators.  For $N=1$, we perform the semi-classical analysis in Appendix \ref{Sec:ParaModuli} and discover that such theories contain non-trivial branches of the moduli space that are isomorphic to certain hyperK\"ahler cones.

\section{A brief review of the 3d supersymmetric index}
\label{Sec:Index}
In this section, we briefly review the 3d supersymmetric index, which we shall refer to as the {\it index} for brievity.  This is the supersymmetric partition function on $S^2 \times  S^1$. It is defined as a trace over states on $S^2 \times \BR$ \cite{Bhattacharya:2008zy,Bhattacharya:2008bja, Kim:2009wb,Imamura:2011su, Kapustin:2011jm, Dimofte:2011py} (we also use the same notation as \cite {Aharony:2013dha, Aharony:2013kma}):
\be
\cI(x, \bm{\mu})\,=\,\Tr \left[ (-1)^{2J_3} x^{\Delta+J_3}\prod_i \mu_i^{T_i} \right]\,,
\ee
where $\Delta$ is the energy in units of the $S^2$ radius (for superconformal field theories, $\Delta$ is related to the conformal dimension), $J_3$ is the Cartan generator of the Lorentz $SO(3)$ isometry of $S^2$, and $T_i$ are charges under non-$R$ global symmetries. The index only receives contributions from the states that satisfy:
\be
\Delta-R-J_3 =  0\,,
\ee
where $R$ is the $R$-charge.
As a partition function on $S^2 \times S^1$, localisation implies that the index receives contributions only from BPS configurations, and it can be written in the following compact way:
\be
\cI(x; \{\bm{\mu},\bm{n}\})\,=\,\sum_{\bm{m}}\frac{1}{|\mathcal{W}_{\bm{m}}|}\int\frac{d\bm{z}}{2\pi i \bm{z}} Z_{\text{cl}}\,Z_{\text{vec}}\,Z_{\text{mat}}\,,
\ee
where we denoted by $\bm{z}$ the fugacities parameterising the maximal torus of the gauge group, and by $\bm{m}$ the corresponding GNO magnetic fluxes on $S^2$. Here $|\mathcal{W}_{\bm{m}}|$ is the dimension of the Weyl group of the residual gauge symmetry in the monopole background labelled by the configuration of magnetic fluxes $\bm{m}$.  We also use $\{\bm{\mu},\bm{n}\}$ to denote possible fugacities and fluxes for the background vector multiplets associated with global symmetries, respectively. 
As usual in localisation computations, the index receives contributions from the non-exact terms of the classical action and from the 1-loop corrections, and each term in the above equation can be described as follows.
\begin{itemize}
\item[$\bm{Z}_{\text{\bf cl}}$:] The classical contribution is associated to Chern-Simons and BF interactions only. Denoting with $k$ the CS level and with $\omega$ and $\mathfrak{n}$ the fugacity and the background flux for the topological symmetry, the classical contribution takes the form
\be
Z_{\text{cl}}\,=\,\prod_{i=1}^{\text{rk}G}\omega^{m_i}z_i^{k\,m_i+\mathfrak{n}}\,,
\ee
where $\text{rk}G$ is the rank of the gauge group $G$.

\item[$\bm{Z}_{\text{\bf vec}}$:] This is the contribution of the $\mathcal{N}=2$ vector multiplet in the theory:
\be
Z_{\text{vec}}\,=\,\prod_{\alpha\in\frak{g}}x^{-\frac{|\alpha(\bm{m})|}{2}}(1-(-1)^{\alpha(\bm{m})}\bm{z}^{\alpha}x^{|\alpha(\bm{m})|})\,
\ee
where $\alpha$ are roots in the gauge algebra $\frak g$.

\item[$\bm{Z}_{\text{\bf mat}}$:] The term encoding the matter fields in the theory enters as 
the product of the contributions of each $\mathcal{N}=2$ chiral field $\chi$, transforming in some representation $\mathcal{R}$ and $\mathcal{R}_F$ of the gauge and the flavour symmetry respectively. Denoting by $r_\chi$ the $R$-charge of $\chi$, its contribution to the index is of the form
\begin{eqnarray}
&&Z_{\chi}\,=\,\prod_{\rho \in \mathcal{R}}\,\prod_{\tilde \rho \in \mathcal{R}_F}\left(\bm{z}^{\rho}\,\bm{\mu}^{\tilde \rho}\,x^{r_{\chi}-1} \right)^{-\frac{|\rho(\bm{m})+\tilde{\rho}(\bm{n})|}{2}}\times\nonumber \\
&&\qquad\qquad\qquad\times\frac{((-1)^{\rho(\bm{m})+\tilde{\rho}(\bm{n})}\,\bm{z}^{-\rho}\,\bm{\mu}^{-\tilde \rho}\,x^{2-r_\chi+|\rho(\bm{m})+\tilde{\rho}(\bm{n})|};x^2)_\infty}{((-1)^{\rho(\bm{m})+\tilde{\rho}(\bm{n})}\,\bm{z}^{\rho}\,\bm{\mu}^{\tilde \rho}\,x^{r_\chi+|\rho(\bm{m})+\tilde{\rho}(\bm{n})|};x^2)_\infty}\,,
\end{eqnarray}
where $\rho$ and $\tilde \rho$ are the weights of $\mathcal{R}$ and $\mathcal{R}_F$ respectively.
\end{itemize}

Let us discuss some examples that will be used later.  The $T(U(1))$ theory is an almost empty theory, containing only the mixed CS coupling between two $U(1)$ background vector multiplets; its index is
\be
\cI_{T(U(1))}(\{{\mu},{n}\},\{{\tau}, {p}\}) =\tau^{n}\mu^{p}~.
\ee
Next, we consider 3d $\CN=4$ $U(1)$ gauge theory with $2$ flavours, whose SCFT is known as $T(SU(2))$.  The index of this theory is
\be \label{indexTSU2}
\begin{split}
&\cI_{T(SU(2))}(\{\bm{\mu},\bm{n}\},\{\bm{\tau}, \bm{p}\})\, \\
&=\sum_{m\in \BZ}\left(\frac{\tau_1}{\tau_2}\right)^{m}\,  \oint\frac{\text{d}z}{2\pi i z}z^{n_1-n_2}\prod_{a=1}^2x^{\frac{|m-p_a|}{2}}\frac{((-1)^{m-p_a}z^{\mp1}\mu_a^{\pm1}x^{3/2+|m-p_a|;x^2})_\infty}{((-1)^{m-p_a}z^{\pm1}\mu_a^{\mp1}x^{1/2+|m-p_a|;x^2})_\infty}\,,
\end{split}
\ee
with the conditions $\mu_1 \mu_2 = \tau_1 \tau_2 =1$ and $n_1+n_2=p_1+p_2=0$ being imposed.

Another important example is the index for $T(U(2))$:
\be
\begin{split}
&\cI_{T(U(2))}(\{\bm{\mu},\bm{n}\},\{\bm{\tau}, \bm{p}\})\, \\
&=\left[ \prod_{i=1}^2 \cI_{T(U(1))}(\{ \mu_i, n_i \}, \{ \tau_i, p_i \})  \right] \times \cI_{T(SU(2))}(\{\bm{\mu},\bm{n}\},\{\bm{\tau}, \bm{p}\})~,
\end{split}
\ee
where in this expression there is no need to impose the constraints on $\{\bm{\mu},\bm{n}\},\{\bm{\tau}, \bm{p}\}$ as for $T(SU(2))$.
Hence we may regard $\{\bm{\mu},\bm{n}\}$ as fugacities and fluxes for the flavour $U(2)$ symmetry, and $\{\bm{\tau}, \bm{p}\}$ as fugacities and fluxes for the enhanced $U(2)$ topological symmetry. The fact that $T(U(2))$ is a self-mirror theory can be translated into the invariance of $\cI_{T(U(2))}(\{\bm{\mu},\bm{n}\},\{\bm{\tau}, \bm{p}\})$ under the simultaneous  exchange $\bm{\mu}\leftrightarrow \bm{\tau}$, $\bm{n}\leftrightarrow \bm{p}$. {For our purpose, we turn off background magnetic fluxes our analyses in the subsequent part of the paper.} 


\subsection{Superconformal multiplets and the index} \label{subsec:multiplets}
Let us now focus on 3d superconformal field theories (SCFTs). The index keeps track of the short multiplets, up to recombination.  This feature makes the reconstruction of the whole content of short multiplets from the index an extremely hard task.
Nevertheless, one may classify the equivalence classes of the multiplets according to their contribution to the index; see \cite{Gadde:2009dj, Beem:2012yn} for 4d SCFTs, and \cite{Razamat:2016gzx} for 3d SCFTs.  For this purpose, it is convenient to set the background magnetic fluxes to zero and expand the index as a power series in $x$
\be
\cI(x,\{\vec \mu, \vec n =0 \})\,=\,\sum_{p=0}^\infty \chi_{p}(\bm{\mu})\,x^{p}\,
\ee
where $\chi_{p}(\bm{\mu})$ is the character of a certain representation of the global symmetry of the theory.  As demonstrated in \cite{Razamat:2016gzx}, one can study the contribution of superconformal multiplets to each order of $x$ in the power series.

Since the shortening conditions for 3d superconformal algebras have been classified \cite{Dolan:2008vc, Cordova:2016emh} (in this paper, we follow the notation of \cite{Cordova:2016emh}), one can extract a lot of useful information about the SCFT in question using the power series of the index.  Indeed this approach has proved successful, in the context of 3d $\CN=2$ gauge theories, for the study of global symmetry enhancement (see \eg~ \cite{Dimofte:2012pd, Razamat:2016gzx, Gang:2018wek, Fazzi:2018rkr}) and supersymmetry enhancement (see \eg~\cite{Evtikhiev:2017heo, Gang:2018huc}).  In this paper, we adopt this approach to study enhancement of supersymmetry and other global symmetries in the context of 3d $S$-fold SCFTs.

As pointed out in \cite{Evtikhiev:2017heo}, it is useful to define the modified index as follows:
\be
\tilde{\CI}(x,\{\vec \mu, \vec n =0 \}) = (1-x^2)\left[ \cI(x,\{\vec \mu, \vec n =0 \})-1 \right]
\ee
Note that all of the terms up to order $x^2$ in the modified index $\tilde{\CI}$ are equal to those in the original index $\CI$ with the same power.  As discussed in \cite{Razamat:2016gzx}, the $\CN=2$ multiplets that can non-trivially contribute to the modified index at order $x^p$ for $p \leq 2$ are as follows:
\be \label{N2multiplets}
\begin{tabular}{|c|c|c|}
\hline
Multiplet & Contribution to the modified index & Comment \\
\hline
$A_2 \bar{B}_1[0]^{(1/2)}_{1/2}$ & $+x^{1/2}$ & free fields  \\
$B_1 \bar{A}_2[0]^{(-1/2)}_{1/2}$ & $-x^{3/2}$ & free fields  \\
$L\bar{B}_1[0]^{(1)}_{1}$ & $+x$ & relevant operators \\
$L\bar{B}_1[0]^{(2)}_{2}$ &$+x^2$ & marginal operators\\
$A_2 \bar{A}_2[0]^{(0)}_1$ & $-x^2$ & conserved currents \\ 
\hline
\end{tabular}
\ee
Indeed, as pointed out in \cite{Razamat:2016gzx, Fazzi:2018rkr} (see also \cite{Beem:2012yn}), the coefficient of $x^2$ in the index counts the number of marginal operators minus the number of conserved currents.

Since our $S$-fold SCFTs has at least $\CN=3$ supersymmetry, we shall work with $\CN=3$ superconformal multiplets.  The ones that are relevant to us are tabulated below, along with the decomposition rules into $\CN=2$ superconformal multiplets \cite{Evtikhiev:2017heo}.
\be \label{N3multiplets}
\begin{tabular}{|l|c|l|}
\hline
Type & $\CN=3$ multiplet & Decomposition into $\CN=2$ multiplets \\
\hline
Flavour current & $B_1[0]^{(2)}_1$ & $L\bar{B}_1[0]^{(1)}_{1} + B_1\bar{L}[0]^{(1)}_{-1}+ A_2 \bar{A}_2[0]^{(0)}_1$ \\
Extra SUSY-current &  $A_2[0]^{(0)}_1$ & $A_2 \bar{A}_2[0]^{(0)}_1 + A_1 \bar{A}_1[1]^{(0)}_{3/2}$ \\
Stress tensor & $A_1[1]^{(0)}_{3/2}$ & $A_1 \bar{A}_1[1]^{(0)}_{3/2}+ A_1 \bar{A}_1[2]^{(0)}_{2}$ \\
\hline
\end{tabular}
\ee

In \cite{Evtikhiev:2017heo}, the author also provided certain conditions on the index regarding supersymmetry enhancement from $\CN=3$.  Let us denoted by $a_p$ the coefficient of $x^p$ in the unrefined modified index $\tilde{\CI}(x,\{\vec \mu=(1, \ldots,1), \vec n =0 \})$, where $\mu_i$ are set to $1$ for all $i$.  A sufficient condition for supersymmetry enhancement states that if $-a_2 > a_1$, then supersymmetry is enhanced from $\CN=3$ to $\CN=3-a_1-a_2$.  The explanation of this condition is as follows. As it can be seen from tables \eref{N2multiplets} and \eref{N3multiplets}, $(-a_2)$ is the number of flavour current multiplets plus the number of extra-SUSY current multiplets, and $a_1$ is the number of flavour current multiplets.  The quantity $(-a_2)-a_1$ is therefore the number of extra-SUSY current multiplets that give rise to the supersymmetry enhancement.  Furthermore, the author of \cite{Evtikhiev:2017heo} also discussed necessary conditions for supersymmetry enhancement to $\CN=4$ and $\CN=5$.  For enhancement to $\CN=4$, one must have $a_1+a_2+2 \geq 0$ and $a_1$ equal to the dimension of the flavour symmetry.  For enhancement to $\CN=5$, one must have $a_1=1$, $a_2 \geq -3$ and $a_p$ even for non-integer $p$.  We emphasise, however, that if one uses the refined index (\ie~ not setting the fugacities $\mu_i$ to unity), one may get more information regarding the presence of the extra SUSY-current multiplets (and hence supersymmetry enhancement), because such a contribution to the index may get cancelled in the unrefined version by the one coming from marginal operators.  We discuss this point in detail in the main text.

\section{A single $U(N)_k$ gauge group with a $T$-link and $n$ flavours} \label{sec:onegaugegroup}
In this section, we consider the following theory:
\be \label{TUNloop}
\begin{tikzpicture}[baseline]
\tikzstyle{every node}=[font=\footnotesize, node distance=0.45cm]
\tikzset{decoration={snake,amplitude=.4mm,segment length=2mm,
                       post length=0mm,pre length=0mm}}
\draw[blue,thick] (0,0) circle (1.5cm) node[midway, right] {$N$ D3};
\draw[decorate,red,thick] (0,1) -- (0,2) node[right] {$J_k= - ST^k$};
\def \n {6}
\def \radius {1.2cm}
\def \margin {0} 
\foreach \s in {1,...,10}
{
	\node[draw=none] (\s) at ({360/\n * (\s - 2)+30}:{\radius-10}) {};
}
\node[draw=none, circle] (last) at ({360/3 * (3 - 1)+30}:{\radius-10}) {};
\node[draw=none,  below right= of 1] (f1) {$\bullet$};
\node[draw=none, above right= of 2] (f2) {};
\node[draw=none, above = of 3] (f3) {};
\node[draw=none, above left= of 4] (f4) {};
\node[draw=none,  below left= of 5] (f5) {$\bullet$};
\node[draw=none,  below = of last] (f6) {$\bullet$};
\node[draw=none] at (-0.9,-1.3) {{\Large $\mathbf{\ddots}$}};
\node[draw=none] at (0,-2.4) {{$n$ D5s}};
\end{tikzpicture}
\qquad \qquad \qquad
\begin{tikzpicture}[baseline]
\tikzstyle{every node}=[font=\footnotesize]
\node[draw, circle] (node1) at (0,1) {$N_k$};
\draw[red,thick] (node1) edge [out=45,in=135,loop,looseness=5, snake it]  (node1);
\node[draw=none] at (1.3,1.5) {{\red $T(U(N))$}};
\node[draw, rectangle] (sqnode) at (0,-1) {$n$};
\draw (node1)--(sqnode);
\end{tikzpicture}
\ee
In \cite{Assel:2018vtq}, the following statements are proposed:
\ben
\item For $k=0$, the SCFT has $\CN=3$ supersymmetry.
\item For $k \geq 3$ and $n=0$, the SCFT has $\CN=4$ supersymmetry.  This statement was confirmed at large $N$ using the corresponding supergravity solutions and the computation of the three sphere partition function in the large $N$ limit.
\een

In the following we compute the superconformal index at low rank $N$ and small values of $n$.  Whenever possible, we deduce the amount of supersymmetry of the SCFT from the index.

\subsection{The abelian case: $N=1$} \label{sec:abelianoneTlink}
The moduli space of this theory was analysed in \cite[sec. 4.4]{Garozzo:2018kra}.  Recall that the $T(U(1))$ is almost an empty theory, with only a prescription for how coupling external gauge fields $A_1$ and $A_2$, which is the supersymmetric completion of the following CS coupling \cite{Gaiotto:2008ak}
\be
-\frac{1}{2 \pi} \int A_1 \wedge dA_2~.
\ee
In \eref{TUNloop}, we identify the $U(1)$ gauge fields $A_1$ and $A_2$ to a single one, and hence the above equation gives rise to a CS level $-2$ to the $U(1)$ gauge group.  In other words, quiver \eref{TUNloop}, with $N=1$, can be identify with the following theory
\be
\begin{tikzpicture}[baseline]
\tikzstyle{every node}=[font=\footnotesize]
\node[draw, circle] (node1) at (0,0) {$1_{k-2}$};
\node[draw, rectangle] (sqnode) at (2,0) {$n$};
\draw (node1)--(sqnode);
\end{tikzpicture}
\ee
where we emphasise that this theory no longer contains a $T$-link. 


As an immediate consequence, for $k=2$, this theory is simply a 3d $\CN=4$ $U(1)$ gauge theory with $n$ flavours.  For $k=2$ and $n=1$, this is dual to a theory of a free hypermultiplet.

Another interesting case is when $k=1$ and $n=1$, which is equivalent to having 3d $\CN=3$ $U(1)_{-1}$ gauge theory with 1 flavour.  The index in this case reads
\be
\begin{split}
\CI_{\eref{TUNloop}, \, N=1, \, k=1, \, n=1}(x ; \omega) &= \CI_{\text{$U(1)_{\pm 1}$ with $1$ flavour}}(x ; \omega) \\
&= 1+x-x^2 \left(\omega + \omega^{-1} +1\right)+x^3 \left(\omega +\omega^{-1}+2\right) \\
& \quad -x^4 \left(\omega+\omega^{-1}+2\right)+x^5 + \ldots~.
\end{split}
\ee
where $\omega$ denotes the topological fugacity. The modified index of this theory is
\be
\begin{split}
&(1-x^2)\left[\CI_{\eref{TUNloop}, \, N=1, \, k=-1}(x ; \omega)-1\right]=x-x^2 \left(\omega + \omega^{-1} +1\right)+\ldots~.
\end{split}
\ee
We expect the enhancement of supersymmetry from $\CN=3$ to $\CN=5$ due to the following argument\footnote{Upon setting $\omega=1$, we obtain the unrefined modified index $x-3x^2+3x^3-x^4-3x^5+\ldots$. Denoting the coefficient of $x^k$ by $a_k$, we see that $(-a_2) = 3 > a_1=1$.  Therefore according to \cite[sec. 4.3]{Evtikhiev:2017heo}, it is expected that supersymmetry gets enhanced from $\CN=3$ to $\CN=3-a_1-a_2=5$.  Moreover, since $a_1=1$, $a_2 =-3 \geq -3$ and $a_p=0$ (which is even) for all non-integers $p$, the necessary condition in \cite[sec. 4.3]{Evtikhiev:2017heo} for having $\CN=5$ supersymmetry is satisfied.}.  The presence of the term $+x$ indicates that there must be an $\CN=3$ flavour current multiplet $B_1[0]^{(2)}_{1}$, which gives rise to the $\CN=2$ multiplet $L\bar{B}_1[0]^{(1)}_1$ contributing $+x$ and the $\CN=2$ multiplet $A_2 \bar{A}_2[0]^{(0)}_1$ contributing $-x^2$.  Since the coefficient of $x^2$ counts the number of marginal operators minus the number of conserved currents \cite{Razamat:2016gzx, Fazzi:2018rkr} (see also \cite{Beem:2012yn}), there must be two extra conserved currents associated with the terms $- \left(\omega+ \omega^{-1} \right)x^2$.  Such extra conserved currents come from two $\CN=3$ extra SUSY-current multiplets $A_2[0]^{(0)}_1$, one carries fugacity $\omega$ and the other carries fugacity $\omega^{-1}$.

\subsection{$U(2)_k$ gauge group and no flavour} \label{sec:U2kwnoflv}
We focus on the following quiver
\be \label{TUNloop2k}
\begin{tikzpicture}[baseline]
\tikzstyle{every node}=[font=\footnotesize]
\node[draw, circle] (node1) at (0,0) {$2_k$};
\draw[red,thick] (node1) edge [out=45,in=135,loop,looseness=5, snake it]  (node1);
\node[draw=none] at (1.3,0.5) {{\red $T(U(2))$}};
\end{tikzpicture}
\ee

We remark that the theory \eref{TUNloop2k} can be also be represented as $T(U(2))/U(2)^{\text{diag}}_{k}$, where the diagonal subgroup $U(2)^\text{diag}$ of the symmetry $U(2) \times U(2)$ of $T(U(2))$ is gauged with CS level $k$.  Nevertheless, we find that the index of such a theory does not depend on the fugacity associated with the topological symmetry, and it is equal to $T(SU(2))/SU(2)^{\text{diag}}_{k}$, where the diagonal subgroup $SU(2)^\text{diag}$ is gauged with CS level $k$.  

In fact, the theory $T(SU(2))/SU(2)^{\text{diag}}_{k}$ was studied in a series of papers \cite{Terashima:2011qi, Gang:2015wya, Gang:2018wek, Gang:2018huc}, mainly in the context of the 3d-3d correspondence.  In particular, it was pointed out in \cite{Gang:2018huc} that for $k=3$, $T(SU(2))/SU(2)^{\text{diag}}_{3}$ is a product of two identical 3d $\CN=4$ SCFTs.  Such an SCFT admits a 3d $\CN=2$ Lagrangian in terms of the $U(1)_{-3/2}$ gauge theory with 1 chiral multiplet carrying gauge charge $+1$ (denoted by $\CT_{-3/2, 1}$), where it turns out that supersymmetry of this theory gets enhanced to $\CN=4$ in the infrared.

In addition to the case of $|k|=3$, we find that the supersymmetry gets enhanced for all $k$ such that $|k| \geq 4$. We summarise the results in the following table.
\begin{longtable}{|c|c|c|c|}
\hline
CS level & Index & Type of $J_k$& Comment \\
\hline
\rowcolor{yellow} $|k|\geq4$ & \eref{TUNloopkleq4} & hyperbolic & \\
\rowcolor{yellow} $|k|=3$ & \eref{loopU2zeroflv} & hyperbolic & Studied in \cite{Gang:2018huc}, a product of two $\CN=4$ SCFTs \\
\hline \hline
$|k|=2$ & diverges & parabolic & \\
\hline \hline
$|k|=1$ & 1 & elliptic& \\
$k=0$ & 1 & elliptic & \\
\hline
\end{longtable}
\noindent We emphasise the cases whose indices indicate supersymmetry enhancement in yellow.  In the following, we discuss the detail of each case.

For $|k| \geq 4$, $J_k$ is hyperbolic. We find that the index reads
\be
\begin{split} \label{TUNloopkleq4}
\CI_{\eref{TUNloop2k}, \, N=2, \, |k|\geq 4}(x) = 1-x^2+2x^3-x^4 + \ldots~.
\end{split}
\ee
where, for each $k$ such that $|k| \geq 4$, the indices differ at order of $x$ greater than $4$.
For example, up to order $x^8$, the indices are as follows:
\be
\begin{array}{ll}
|k|=4 &\qquad  1 - x^2 + 2 x^3 - x^4 - 4 x^5 + 10 x^6-10 x^7+8 x^8+ \ldots \\
|k|=5 &\qquad 1 - x^2 + 2 x^3 - x^4 - 2 x^5 + 6 x^6-8 x^7 +4 x^8 +\ldots \\
|k|=6 &\qquad  1 - x^2 + 2 x^3 - x^4 - 2 x^5 + 6 x^6-8 x^7+6 x^8+ \ldots
\end{array}
\ee
The modified index is
\be
(1-x^2)\left[\CI_{\eref{TUNloop2k}, \, N=2, \, |k|\geq 4}(x) -1\right] = -x^2+2x^3+\ldots~.
\ee
The fact that the coefficient of $x$ vanishes implies that we have no $\CN=3$ flavour current multiplet $B_1[0]^{(2)}_{1}$.  
The term $-x^2$ indicates the presence of the $\CN=3$ extra SUSY-current multiplet $A_2[0]^{(0)}_1$. We thus conclude that the supersymmetry gets enhanced from $\CN=3$ to $\CN=4$ when $|k| \geq 4$.  

For $|k|=3$, the index reads
\be \label{loopU2zeroflv}
\CI_{\eref{TUNloop2k}, \, N=2, \, |k|=3}(x) =  1 - 2 x^2 + 4 x^3 - 3 x^4 + \ldots~.
\ee 
According to \cite{Gang:2018huc}, this is equal to the square of the index of $\CT_{-3/2, 1}$.  In the notation adopted by this paper, the index of $\CT_{-3/2, 1}$ reads
\be
\CI_{\CT_{-3/2,1}} (x; w) = 1-x^2 + \left(w +w^{-1}  \right) x^3 - 2x^4 + \ldots~.
\ee
where $w$ is the topological fugacity.  Indeed, we find that
\be
\left[ \CI_{\CT_{-3/2,1}} (x; w=1) \right]^2 =  \CI_{\eref{TUNloop2k}, \, N=2, \, k=-3}(x) ~.
\ee
The modified index corresponding to \eref{loopU2zeroflv} reads
\be \label{modifiedN2n0}
(1-x^2)\left[\CI_{\eref{TUNloop2k}, \, N=2, \, |k|=3}(x) -1\right] = -2 x^2 + 4 x^3 - x^4 \ldots~.
\ee
Let us denote the coefficient of $x^p$ by $a_p$. Naively, from the condition $-a_2 = 2 > a_1=0$ discussed in \cite{Evtikhiev:2017heo}, one might expect that supersymmetry gets enhanced to $\CN=3-a_1-a_2=5$.  However, this cannot be true, for the reason that the $\CN=5$ stress tensor multiplet in the representation $[1,0]$ of $SO(5)$ decomposes into one $\CN=2$ multiplet $L\bar{B}_1[0]^{(1)}_1$, which contributes $a_1=1$ \cite[(B.25)]{Evtikhiev:2017heo} (but here we have $a_1=0$).  Since this theory is a product of two copies of $\CT_{-3/2, 1}$, which has enhanced $\CN=4$ supersymmetry, there are two copies of the $\CN=3$ extra SUSY-current multiplet $A_2[0]^{(0)}_{\Delta=1}$.  This is consistent with the fact that the modified index has $a_1=0$ and $a_2=-2$.


For the theory with $|k|=2$ ($J_{2}$ is parabolic), the index diverges, and so we have a ``bad'' theory in the sense of \cite{Gaiotto:2008ak}.  For $|k|=1$ and $k=0$ ($J_k$ is elliptic in these cases), we find that the index is equal to unity.  

\subsection{Adding one flavour ($n=1$) to the $U(2)_k$ gauge group} \label{sec:oneflvU2k}
We now consider the following theory
\be \label{TUNloop2kn1}
\begin{tikzpicture}[baseline]
\tikzstyle{every node}=[font=\footnotesize]
\node[draw, circle] (node1) at (0,0) {$2_k$};
\draw[red,thick] (node1) edge [out=45,in=135,loop,looseness=5, snake it]  (node1);
\node[draw=none] at (1.3,0.5) {{\red $T(U(2))$}};
\node[draw, rectangle] (sqnode) at (3,0) {$1$};
\draw (node1)--(sqnode);
\end{tikzpicture}
\ee
Let us summarise the results in the following table.
\begin{longtable}{|c|c|c|c|}
\hline
CS level & Index & Type of $J_k$& Comment \\
\hline
$k=-2$ & \eref{oneflvkmm2} & parabolic & \\
\hline \hline
$k=-1$ & \eref{oneflvotherk} & elliptic & \\
$k=0$ & \eref{oneflvotherk} & elliptic & \\
\rowcolor{yellow} $k=1$ & \eref{oneflvkm1} & elliptic & \\
\hline \hline
\rowcolor{yellow} $k=2$ & \eref{oneflvkm2} & parabolic & A free hyper $\times$ an $\CN=4$ SCFT \\
\hline
$|k| \geq 3$ & \eref{oneflvotherk} & hyperbolic & \\
\hline
\end{longtable}
\noindent where we emphasise the cases that have supersymmetry enhancement in yellow.

For $k=1$, we find that the index reads
\be \label{oneflvkm1}
\begin{split}
\CI_{\eref{TUNloop2kn1}, \, k=1}(x; \omega) &= 1+x+x^2 \left[ 1-(1+\omega + \omega^{-1})\right] -x^3 \left(\omega +\omega^{-1} \right)\\
& \quad +x^4(4+\omega^2 + \omega^{-2}  + 3 \omega+ 3 \omega^{-1})+ \ldots~,
\end{split}
\ee
where $\omega$ is the topological fugacity.  From the above expression, we find that the modified index is as follows:
\be
(1-x^2) \left[ \CI_{\eref{TUNloop2kn1}, k=1}(x; \omega)-1 \right] = x+x^2 \left[ 1-(1+\omega + \omega^{-1})\right]+\ldots~.
\ee
From this, one can see the enhancement of supersymmetry from $\CN=3$ to $\CN=5$ as follows. The presence of the term $+x$ indicates that there must be an $\CN=3$ flavour current multiplet $B_1[0]^{(2)}_{1}$, which gives rise to the $\CN=2$ multiplet $L\bar{B}_1[0]^{(1)}_1$ contributing $+x$ and the $\CN=2$ multiplet $A_2 \bar{A}_2[0]^{(0)}_1$ contributing $-x^2$.  Since the coefficient of $x^2$ counts the number of marginal operators minus the number of conserved currents \cite{Razamat:2016gzx, Fazzi:2018rkr} (see also \cite{Beem:2012yn}), there must be an $\CN=2$ marginal operator (in the multiplet $L\bar{B}_1[0]^{(2)}_2$) contributing $+x^2$ to cancel the aforementioned contribution $-x^2$, and there must be two extra conserved currents associated with the terms $- \left(\omega+ \omega^{-1} \right)x^2$.  The latter can only come from two copies of the $\CN=3$ extra SUSY-current multiplet $A_2[0]^{(0)}_1$, carrying the global symmetry associated with $\omega$ and $\omega^{-1}$. (This gives rise to two copies of $\CN=2$ $A_2 \bar{A}_2[0]^{(0)}_1$ multiplet contributing the term $- \left(\omega+ \omega^{-1} \right)x^2$.)  The presence of such a multiplet leads to the enhancement of supersymmetry from $\CN=3$ to $\CN=5$\footnote{We remark that one has to use the sufficient condition stated in \cite[sec. 4.3]{Evtikhiev:2017heo} with great care.  Upon setting $\omega=1$ in the modified index, we obtain $x-2x^2$.  Denoting the coefficient of $x^k$ by $a_k$, we see that $-a_2 = 2 > a_1=1$, and from \cite{Evtikhiev:2017heo}, one might naively expect that supersymmetry gets enhanced to $\CN=3-a_1-a_2=4$, because we have only $(-a_2)-a_1=1$ extra SUSY-current multiplet.  The unrefinement of the index is misleading here, because we in fact have two extra SUSY-current multiplets carrying the global fugacities $\omega$ and $\omega^{-1}$, and these cannot be cancelled with $-1$ at order $x^2$ in the index.  The reason for us to write $x^2 \left[ 1-(1+\omega + \omega^{-1})\right]$ is to show explicitly that the contribution $-1$ of the conserved current has to be cancelled with the contribution $+1$ from the marginal operator, which is neutral under the symmetry associated with $\omega$. Note that since $a_1=1$, $a_2 =-2 \geq -3$ and $a_p=0$ (which is even) for all non-integers $p$, the necessary condition in \cite[sec. 4.3]{Evtikhiev:2017heo} for having $\CN=5$ supersymmetry is satisfied.}.

For $k=2$, the index reads
\be \label{oneflvkm2}
\begin{split}
&\CI_{\eref{TUNloop2kn1},\,  k=2}(x; w) \\
&=1+\left(w+\frac{1}{w}\right) x^{\frac{1}{2}}+\left(2 w^2+\frac{2}{w^2}+2\right) x+\left( 2 w^3+\frac{2}{w^3}+2 w+\frac{2}{w} \right)x^{\frac{3}{2}}\\
&\quad+\left(3 w^4+\frac{3}{w^4}+2 w^2+\frac{2}{w^2}+1\right) x^2 + \ldots
\end{split}
\ee
The term $x^{1/2}$ indicates that this theory contains a free part due to the fact that the $R$-charge of the basic monopole operators hits the unitary bound.  The above index can be rewritten as
\be \label{freeandintSCFT}
\CI_{\eref{TUNloop2kn1},\,  k=2}(x; w)= \CI_{\text{free}}(x; w)  \times \CI^{\eref{TUNloop2kn1},\,  k=2}_{\text{SCFT}}(x;w) 
\ee
where the index of a free hypermultiplet is given by
\be \label{freeindex}
\CI_{\text{free}}(x; w) =  \frac{(x^{2-\frac{1}{2}} w ;x^2)_\infty}{(x^{\frac{1}{2}} w^{-1} ;x^2)_\infty}  \frac{(x^{2-\frac{1}{2}} w^{-1} ;x^2)_\infty}{(x^{\frac{1}{2}} w ;x^2)_\infty}
\ee
and the index of the interacting SCFT part is
\be\label{indexintSCFTn1}
\begin{split}
&\CI^{\eref{TUNloop2kn1},\,  k=2}_{\text{SCFT}}(x; w)  \\
&= 1+x \left(w ^2+\frac{1}{w ^2}+1\right) +x^2 \left(w ^4+\frac{1}{w ^4}-1\right) +x^{5/2} \left(-w -\frac{1}{w }\right)+\ldots \\
&=1+x \chi^{SU(2)}_{[2]}(w) +x^2 \left[ \chi^{SU(2)}_{[4]}(w) - (\chi^{SU(2)}_{[2]}(w) +\chi^{SU(2)}_{[0]}(w) ) \right] \\
& \qquad -x^{\frac{5}{2}} \chi^{SU(2)}_{[2]}(w) +\ldots~,
\end{split}
\ee
with the unrefinement
\be
\CI^{\eref{TUNloop2kn1},\,  k=2}_{\text{SCFT}}(x; w=1)  = 1 + 3 x + x^2 - 2 x^{5/2} + 4 x^3 + 4 x^{7/2} + 3 x^4 + \ldots~.
\ee
As can be seen from \eref{indexintSCFTn1}, the interacting SCFT has enhanced $\CN=4$ supersymmetry.  The argument is similar to the one used before.  The term $+x \chi^{SU(2)}_{[2]}(w)$ indicates that the theory has an $SU(2)$ flavour symmetry. Indeed, there is an $\CN=3$ flavour current multiplet $B_1[0]^{(2)}_{1}$ transforming in the adjoint representation $[2]$ of this symmetry; this gives rise to the $\CN=2$ multiplet $L\bar{B}_1[0]^{(1)}_1$ contributing $+x \chi^{SU(2)}_{[2]}(w)$ and the $\CN=2$ multiplet $A_2 \bar{A}_2[0]^{(0)}_1$ contributing $-x^2 \chi^{SU(2)}_{[2]}(w)$.  The term $+x^2\chi^{SU(2)}_{[4]}(w)$ corresponds to the $\CN=2$ marginal operator\footnote{It should be noted that the 2nd symmetric power of $[2]$ is $\Sym^2[2] = [4]+[0]$. The representation $[4]$, appearing at order $x^2$ of the index, is a part of this symmetric power. \label{footnotesinglet}} in the multiplet $L\bar{B}_1[0]^{(2)}_2$.  It can be clearly seen that there is another conserved current corresponding to the term $-x^2 \chi^{SU(2)}_{[0]}(w)$.  Indeed, the latter comes from the $\CN=3$ extra SUSY-current multiplet $A_2[0]^{(0)}_1$ in the trivial representation $[0]$ of $SU(2)$; this gives rise to an $\CN=2$ conserved current multiplet $A_2 \bar{A}_2[0]^{(0)}_1$ contributing the term $-x^2 \chi^{SU(2)}_{[0]}(w)$.  The existence of the extra SUSY-current multiplet indicates that there is an enhancement of supersymmetry from $\CN=3$ to $\CN=4$.

For $k=0, \, -1$ and $|k|\geq 3$, we find that the index reads
\be \label{oneflvotherk}
1+x+0x^2 +\ldots~.
\ee
The term $+x$ indicates that there must be an $\CN=3$ flavour current multiplet $B_1[0]^{(2)}_{1}$, which gives rise to the $\CN=2$ multiplet $L\bar{B}_1[0]^{(1)}_1$ contributing $+x$ and the $\CN=2$ multiplet $A_2 \bar{A}_2[0]^{(0)}_1$ contributing $-x^2$.  Hence the theory has a $U(1)$ flavour symmetry. The fact that the term $x^2$ vanishes implies that there is an $\CN=2$ marginal operator in the multiplet $L\bar{B}_1[0]^{(2)}_2$, contributing $+x^2$, which cancels the aforementioned $-x^2$ term.  Hence, in this case, there is no signal of the existence of the extra SUSY-current multiplet, \ie~ we cannot deduce the enhancement of supersymmetry. 

For $k=-2$, we find that the index reads
\be \label{oneflvkmm2}
1 + 2 x + x^2 + 8 x^4+\ldots~.
\ee
There are two $\CN=3$ flavour current multiplet $B_1[0]^{(2)}_{1}$ which gives rise to two copies of $\CN=2$ multiplets $L\bar{B}_1[0]^{(1)}_1$ contributing $+2x$ and two copies of $\CN=2$ multiplets $A_2 \bar{A}_2[0]^{(0)}_1$ contributing $-2x^2$.  Hence the theory has a $U(1)^2$ flavour symmetry.  We may construct three $\CN=2$ marginal operators by taking a symmetric product of two relevant operators in the $L\bar{B}_1[0]^{(1)}_1$ multiplets.  Their contribution $+3x^2$ cancels the aforementioned $-2x^2$ and yields $+x^2$.  There is no signal of the existence of the extra SUSY-current multiplet, \ie~ we cannot deduce the enhancement of supersymmetry.

\subsection{Adding $n$ flavours to the $U(2)_2$ gauge group} \label{sec:addinggenflv}
In this section, we add an arbitrary number of flavours to the parabolic case\footnote{Here $J_2=-ST^2$ is a parabolic element of $SL(2,\BZ)$.  It is related to $T^{-1}$ by the following similarity transformation: $(TST) J_2 (TST)^{-1} =T^{-1}$.  However, we emphasise that, when fundamental flavours are added as in \eref{TUNloop2kngeq2}, the theory is different from $U(2)_{-1}$ with $n$ flavours.  This can be seen clearly from the indices.  For example, for $n=1$, the index for $U(2)_{-1}$ with 1 flavour is $1$ but \eref{freeandintSCFT} is non-trivial. \label{nontrivial}}, namely 
\be \label{TUNloop2kngeq2}
\begin{tikzpicture}[baseline]
\tikzstyle{every node}=[font=\footnotesize]
\node[draw, circle] (node1) at (0,0) {$2_2$};
\draw[red,thick] (node1) edge [out=45,in=135,loop,looseness=5, snake it]  (node1);
\node[draw=none] at (1.3,0.5) {{\red $T(U(2))$}};
\node[draw, rectangle] (sqnode) at (3,0) {$n$};
\draw (node1)--(sqnode);
\end{tikzpicture}
\ee
When the number of flavours is one (\ie ~ $n=1$), we have seen from \eref{freeandintSCFT} that the theory factorises into a product of the theory of a free hypermultiplet and an interacting SCFT with enhanced $\CN=4$ supersymmetry.  For $n \geq 2$, the index does not exhibit explicitly the presence of the extra SUSY-current multiplet.  Nevertheless, as we demonstrate below, the theory still has interesting physics that is bares certain resemblance to the 3d $\CN=4$ $U(1)$ gauge theory with $n$ flavours, such as the properties of monopole operators.



For concreteness, let us first consider the case of $n=2$. The index reads\footnote{The symmetric product of the representation $[2;0]+[0;2]$ of $SU(2) \times SU(2)$ is $2[0;0]+[4;0]+[0;4]+[2;2]$. The representation in the first bracket of order $x^2$ (\ie~ those with plus signs) can be written as $\Sym^2 ([2;0]+[0;2]) + [4;0]+[0;2]-[0;0]$. In the same way as in footnote \ref{footnotesinglet}, one singlet in the decomposition of the symmetric power does not participate in the index; this explains the term $-[0;0]$.  Moreover, it is worth pointing out that, in this case, there are extra representations that are not contained in the symmetric product, namely $[4;0]$ and $[0;2]$.}
\be \label{U22twoflavours}
\begin{split}
&\CI_{\eref{TUNloop2kngeq2}, \, n=2} (x; \omega, y) \\
&=1+ x \left[ \chi^{SU(2)}_{[2]}(\omega) + \chi^{SU(2)}_{[2]}(y)   \right] +  x^2 \Big[\Big(1+ 2\chi^{SU(2)}_{[4]}(\omega) + \chi^{SU(2)}_{[4]}(y)   \\
& \quad +\chi^{SU(2)}_{[2]}(\omega) \chi^{SU(2)}_{[2]}(y) +\chi^{SU(2)}_{[2]}(y) \Big) - \left( \chi^{SU(2)}_{[2]}(\omega) + \chi^{SU(2)}_{[2]}(y)   \right)  \Big] +\ldots~,
\end{split}
\ee
with the unrefinement
\be
\CI_{\eref{TUNloop2kngeq2}, \, n=2} (x; \omega=1, y=1)=1 + 6 x + 22 x^2 + 18 x^3 + 29 x^4+ \ldots~.
\ee
where the topological fugacity is denoted by $w=\omega^2$.  We see that the $U(1)$ topological symmetry gets enhanced to $SU(2)$.  This phenomenon also occurs for 3d $\CN=4$ $U(1)$ gauge theory with $2$ flavours, whose index is
\be
\begin{split}
 &\CI_{T(SU(2))}(x; \omega, y)  \\
 &= 1+ x \left[ \chi^{SU(2)}_{[2]}(\omega) + \chi^{SU(2)}_{[2]}(y)   \right] \\
 & \qquad +  x^2 \left[  \chi^{SU(2)}_{[4]}(\omega) + \chi^{SU(2)}_{[4]}(y)  -  \left( \chi^{SU(2)}_{[2]}(\omega) + \chi^{SU(2)}_{[2]}(y) +1 \right) \right] + \ldots
 \end{split}
\ee
with the unrefinement
\be
\CI_{T(SU(2))}(x; \omega=1, y=1) = 1 + 6 x + 3 x^2 + 6 x^3 + 17 x^4+\ldots~,
\ee

For $n=3$, we find that the index of \eref{TUNloop2kngeq2} reads
\be \label{TUNloop2kn3}
\begin{split}
&\CI_{\eref{TUNloop2kngeq2}, \, n=3} (x; w, y)  \\
&= 1+ x \left[ 1+\chi^{SU(3)}_{[1,1]}(\vec y) \right] + x^{\frac{3}{2}} ( w+ w^{-1}) \\
& \qquad + x^2 \Big[ \chi^{SU(3)}_{[2,2]}(y) +2\chi^{SU(3)}_{[1,1]}(\vec y)+1 \Big]+ \ldots~,
\end{split}
\ee
with the unrefinement
\be
\begin{split}
\CI_{\eref{TUNloop2kngeq2}, \, n=3} (x; w=1, \vec{y}=(1,1)) &= 1 + 9 x + 2 x^{\frac{3}{2}} + 44 x^2 + 18 x^{\frac{5}{2}} \\
& \quad + 117 x^3 + 34 x^{\frac{7}{2}} + 188 x^4+\ldots~,
\end{split}
\ee
where $w$ the topological fugacity and $\vec y$ the $SU(3)$ flavour fugacities.  Again, this bares some similarity with the $U(1)$ gauge theory with $3$ flavours, whose index is
\be
\begin{split}
\CI_{T_{(2,1)}(SU(3))}(x; w, \vec y) &= 1+ x \left[ 1+\chi^{SU(3)}_{[1,1]}(\vec y) \right] +x^{\frac{3}{2}}  (w+ w^{-1})  \\
& \qquad - x^2 \left[  \chi^{SU(3)}_{[2,2]}(\vec y) -(1+\chi^{SU(3)}_{[1,1]}(\vec y)) \right] + \ldots~,
\end{split}
\ee
with the unrefinement
\be
\CI_{T_{(2,1)}(SU(3))}(x; w=1, \vec y=(1,1)) =  1 + 9 x + 2 x^{\frac{3}{2}} + 18 x^2 + 21 x^3 + 54 x^4+ \ldots~,
\ee

We observe that for a general $n$, \eref{TUNloop2kngeq2} has a global symmetry $SU(n) \times U(1)$, where the $U(1)$ is the topological symmetry, which is enhanced to $SU(2)$ for $n=2$.  Moreover, the terms $x^{\frac{n}{2}} (w+ w^{-1})$ indicate that theory \eref{TUNloop2kngeq2} contains the basic monopole operators $V_{\pm(1,0)}$ (with flux $\pm(1,0)$ under the $U(2)$ gauge group), carrying $R$-charge $\frac{n}{2}$, similar to $V_{\pm1}$ in the $U(1)$ gauge theory with $n$ flavours. Moreover, with CS level $k=2$, these basic monopole operators are gauge neutral, so they are gauge invariant themselves without any dressing by a chiral field in the fundamental hypermultiplet\footnote{Note that this statement does not hold when the CS level is not equal to $2$, and in order to form a gauge invariant combination, the monopole operators need to be dressed by chiral fields in the fundamental hypermultiplet.}. A non-trivial physical implication is that the contribution of the $T$-link cancel the contribution of the non-abelian vector multiplet in the $R$-charge of the monopole operator.

\subsection{Duality with theories with two gauge groups}
Here we examine the duality between the following theories
\be \label{twodualtheories}
\scalebox{0.8}{
\begin{tikzpicture}[baseline]
\tikzstyle{every node}=[font=\footnotesize]
\node[draw, circle] (node1) at (0,0) {$N_k$};
\draw[red,thick] (node1) edge [out=45,in=135,loop,looseness=5, snake it]  (node1);
\node[draw=none] at (0,1.4) {{\red $T(U(N))$}};
\node[draw, rectangle] (sqnode) at (2,0) {$n$};
\draw (node1)--(sqnode);
\end{tikzpicture}}
\qquad \qquad \qquad 
\scalebox{0.8}{
\begin{tikzpicture}[baseline]
\tikzstyle{every node}=[font=\footnotesize]
\node[draw, circle] (node1) at (-1.5,0) {$N_{0}$};
\node[draw, circle] (node2) at (1.5,0) {$N_{k}$};
\node[draw, rectangle] (sqnode2) at (3.5,0) {$n-1$};
\draw[draw=black,solid,thick,-]  (node1) to[bend right=20]   (node2) ; 
\draw[draw=black,solid,thick,-]  (node2)--(sqnode2) ; 
\draw[draw=red,solid,thick,-,snake it]  (node1) to[bend left=20]  node[midway,above] {{\red $T(U(N))$}}  (node2) ; 
\end{tikzpicture}}
\ee
This duality can be seen from the brane system by moving one of the D5-brane across the $J$-fold and, thereby, turning it into an NS5 brane.  For general values of $N$ and $k$, both theories have a global symmetry $U(n)$.  However, as can be seen from the indices, they arise from different origins in the quiver description.  

Let us take, for example, $N=2$, $k=2$ and $n=3$.  The index of the left quiver is given by \eref{TUNloop2kn3}.
The index of the right quiver reads
\be \label{twogaugenodeindexn3}
\begin{split}
&1+x \left[ 2+ (w_1 + w_1^{-1}) \chi^{SU(2)}_{[1]} (\tilde{y}) +\chi^{SU(2)}_{[2]}(\tilde{y})\right]+x^{\frac{3}{2}} \left(w_1 w_2+\frac{1}{w_1 w_2}\right)  \\
& \quad + x^2 \Big[4 + (w_1^2 +3 + w_1^{-2}) \chi^{SU(2)}_{[2]} (\tilde{y}) + (w_1 + w_1^{-1}) \left( \chi^{SU(2)}_{[3]} (\tilde{y})+2\chi^{SU(2)}_{[1]} (\tilde{y}) \right)\\
& \quad + \chi^{SU(2)}_{[4]} (\tilde{y})  \Big]+ \ldots~.
\end{split}
\ee
where $w_1$ and $w_2$ are the topological fugacities associated with the left and right nodes, and we denote the flavour fugacities by $\tilde{y}$.
This expression can be rewritten in the way that the $SU(3)$ symmetry is manifest by setting
\be
w_1 = y_1^{-\frac{3}{2}}~, \qquad \tilde{y} = y_1^{-\frac{1}{2}} y_2 ~,
\ee
upon which we recover the expression \eref{TUNloop2kn3}.

From the coefficient of $x$, we see that the mesons in the adjoint representation $[1,1]$ of $SU(3)$ of the left quiver in \eref{twodualtheories} are mapped to the following operators of the right quiver in \eref{twodualtheories}: 
\ben
\item the mesons in the adjoint representation $[2]$ of the $SU(2)$ flavour symmetry; 
\item the dressed monopole operators in the fundamental representation $[1]$ of $SU(2)$ and carrying topological charges $\pm 1$ under the left node\footnote{This is similar to the dressed monopole operators \eref{dressedmon0kabel} in the abelian theory.}; and 
\item the trace of the adjoint chiral field associated with the left node.
\een
Moreover, by comparing the terms at order $x^{\frac{3}{2}}$ in \eref{TUNloop2kn3} and \eref{twogaugenodeindexn3}, we see that the basic monopole operators $V_{\pm}$, carrying topological charges $\pm 1$, in the left quivers are mapped to the basic monopole operators $V_{\pm(1,1)}$, carrying topological charges $\pm (1,1)$, in the right quivers.  

These statements can be easily generalised to other values of $k$ and $n$.   


\section{$U(2)_{k_1} \times U(2)_{k_2}$ with two $T$-links} \label{sec:twoTtwogaugegroups}
In this subsection we consider the following theory
\be \label{twoTlinksnoflv}
\scalebox{0.8}{
\begin{tikzpicture}[baseline]
\tikzstyle{every node}=[font=\footnotesize]
\draw[blue,thick] (0,0) circle (1.5cm) node[midway, right] {$2$ D3};
\tikzset{decoration={snake,amplitude=.4mm,segment length=2mm,
                       post length=0mm,pre length=0mm}}
\draw[decorate,red,thick] (0,1) -- (0,2) node[right] {$J_{k_1}=-ST^{k_1}$};
\draw[decorate,red,thick] (0,-1)--(0,-2) node[right] {$J_{k_2}=-ST^{k_2}$};
\end{tikzpicture}}
\qquad\qquad\qquad
\scalebox{0.8}{
\begin{tikzpicture}[baseline]
\tikzstyle{every node}=[font=\footnotesize]
\node[draw, circle] (node1) at (-1.5,0) {$2_{k_1}$};
\node[draw, circle] (node2) at (1.5,0) {$2_{k_2}$};
\draw[draw=red,solid,thick,-,snake it]  (node1) to[bend right=20]  node[midway,below] {{\red $T(U(2))$}}  (node2) ; 
\draw[draw=red,solid,thick,-,snake it]  (node1) to[bend left=20]  node[midway,above] {{\red $T(U(2))$}}  (node2) ; 
\end{tikzpicture}}
\ee

The three sphere partition function as well as the supergravity solution corresponding to $U(N)_{k_1} \times U(N)_{k_2}$ gauge group (\ie~ $N$ D3 branes), in the large $N$ limit, were studied in \cite{Assel:2018vtq}.  In such a reference, the CS levels were restricted such that $\tr(\pm J_{k_1} J_{k_2}) >2$, equivalently $\pm(k_1 k_2 -2) >2$, where the sign $\pm$ is chosen such that the trace is greater than 2.   In which case, $J_{k_1} J_{k_2}$ is a hyperbolic element of $SL(2,\BZ)$, and the theory was predicted to have $\CN=4$ supersymmetry in the large $N$ limit.   Here, instead, we focus on the superconformal indices and supersymmetry enhancement when the gauge group is taken to be $U(2)_{k_1} \times U(2)_{k_2}$ for general values of $k_1$ and $k_2$.

Note that if one of $k_1$ or $k_2$ is $1$, say $k_1=1$, we have $J_{1} J_{k_2}=ST ST^{k_2}$.  This is related by a $T$-similarity transformation to $TJ_1 J_{k_2} T^{-1} = T(STSTT^{k_2-1})T^{-1}= -ST^{k_2-2} = J_{k_2-2}$, where have used the identity $TSTST=-S$ (see also \cite[footnote 19]{Assel:2018vtq}).  In other words, the two duality walls $J_{1}$ and $J_{k_2}$ can be reduced to a single duality wall $J_{k_2-2}$ (assuming that there are no NS5 and D5 branes).  Henceforth, we shall not consider such a possibility in the absence of hypermultiplet matter.


In general, we observe that whenever $J_{k_1} J_{k_2}$ is a {\it parabolic} element of $SL(2,\BZ)$, \ie~ $|\tr (J_{k_1} J_{k_2})|=|k_1k_2-2|=2$ or equivalently $k_1k_2=0$ or $4$, the index diverges and the theory is ``bad'' in the sense of \cite{Gaiotto:2008ak}.  In which case, we cannot deduce the low energy behaviour of the theory from its quiver description. 

%

We observe that the index of \eref{twoTlinksnoflv} does not depend on the fugacities associated with the topological symmetries.  Similarly to section \ref{sec:U2kwnoflv}, the gauge group in \eref{twoTlinksnoflv} can be taken to be $SU(2)_{k_1} \times SU(2)_{k_2}$ and this yields the same index.

Let us now take $k_1=2$ and examine various values of $k_2$ as follows.
\begin{longtable}{|c|c|c|c|}
\hline
CS levels $(k_1, k_2)$ & Index & Type of $J_{k_1} J_{k_2}$ & Comment \\
\hline
$(2,5)$  & $1+x^4+\ldots$   & hyperbolic & \\
$(2,4)$  & $1+x^4+\ldots$  & hyperbolic  & \\
\rowcolor{yellow} \bf $(2,3)$  &  \bf $1-x^2+2 x^3-x^4+\ldots$ & hyperbolic&  New, SUSY enhancement \\
\hline\hline
$(2,2)$  & diverges  & parabolic &  \\
\hline\hline
$(2,1)$  & $1$ &  elliptic & Same as \eref{TUNloop2k}, $k=0$  \\
\hline\hline
$(2,0)$  & diverges  & parabolic &  \\
\hline\hline
\rowcolor{yellow}  $(2,-1)$  & \eref{TUNloopkleq4}   &  hyperbolic &Same as \eref{TUNloop2k}, $k=\pm4$ \\
$(2,-2)$  & $1+x^4+\ldots$   & hyperbolic &  \\
$(2,-3)$  & $1+x^4+\ldots$   & hyperbolic & \\
$(2,-4)$  & $1+x^4+\ldots$  & hyperbolic  & \\
\hline
\end{longtable}
\noindent The cases whose indices exhibit supersymmetry enhancement are emphasised in yellow. The indices for the cases not highlighted in yellow do not signalise the presence of extra SUSY-current multiplets.  The CS levels $(k_1, k_2)=(2,3)$ gives a new SCFT with enhanced $\CN=4$ supersymetry, whereas the case with $(k_1, k_2)=(2,-1)$ is the same as theory \eref{TUNloop2k} with $k=-4$, which also has supersymmetry enhancement to $\CN=4$.

For $k_1=3$, we find a similar pattern, as tabulated below.  Unfortunately, the cases that has supersymmetry enhancement, namely $(k_1,k_2)=(3,2)$ and $(3,-1)$, are identical with certain theories that have been discussed before.
\begin{longtable}{|c|c|c|c|}
\hline
CS levels $(k_1, k_2)$ & Index & Type of $J_{k_1} J_{k_2}$ & Comment \\
\hline
$(3,4)$  & $1+x^4+\ldots$  & hyperbolic & \\
$(3,3)$  & $1+2x^4+\ldots$  & hyperbolic &  \\
\rowcolor{yellow} $(3,2)$  & $1-x^2+2x^3+\ldots$ & hyperbolic & Same as $(k_1,k_2)=(2,3)$ \\
\hline \hline
 $(3,1)$ & {$1$}  & elliptic & Same as \eref{TUNloop2k}, $k=\pm1 $  \\
\hline\hline
$(3,0)$  & diverges  & parabolic  & \\
\hline \hline
\rowcolor{yellow}  $(3,-1)$ & \eref{TUNloopkleq4}  &  hyperbolic &  Same as \eref{TUNloop2k}, $k=\pm5$  \\
$(3,-2)$  & $1+x^4+\ldots$  & hyperbolic &  \\
\hline
\end{longtable}

\subsection{Adding flavours to the parabolic case} \label{subsec:addflvtwoTtwonodes}
In this section, we add fundamental flavours to either or both nodes in the parabolic case.  For definiteness, we consider the theory involving two $J_2$ duality walls\footnote{Similarly to the remark in footnote \ref{nontrivial}, even though $J_2^2$ is related to $T^{-2}$ by a similarity transformation in $SL(2,\BZ)$, upon adding hypermultiplet matter, the theory becomes non-trivial.} and a collection of D5 branes arranged in the following way:
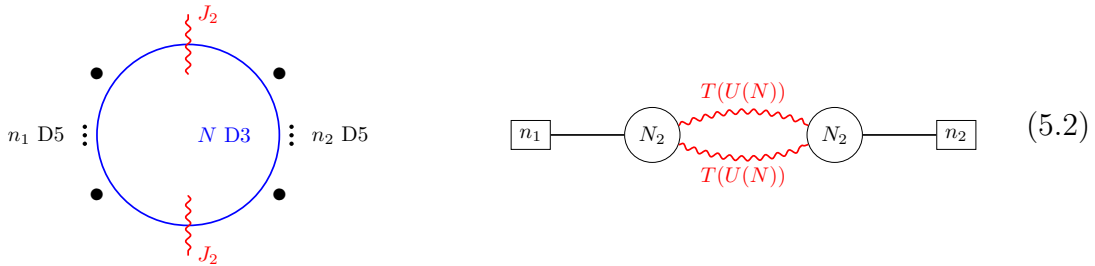
\be \label{flvpara}
\scalebox{0.8}{
\begin{tikzpicture}[baseline]
\tikzstyle{every node}=[font=\footnotesize]
\draw[blue,thick] (0,0) circle (1.5cm) node[midway, right] {$N$ D3};
\node[draw=none] at (-1.5,1) {\large{$\bullet$}};
\node[draw=none] at (-1.7,0.1) {\Huge{$\vdots$}};
\node[draw=none] at (-2.5,0) {$n_1$ D5};
\node[draw=none] at (-1.5,-1) {\large{$\bullet$}};
\node[draw=none] at (1.5,1) {\large{$\bullet$}};
\node[draw=none] at (1.7,0.1) {\Huge{$\vdots$}};
\node[draw=none] at (2.5,0) {$n_2$ D5};
\node[draw=none] at (1.5,-1) {\large{$\bullet$}};
\tikzset{decoration={snake,amplitude=.4mm,segment length=2mm,
                       post length=0mm,pre length=0mm}}
\draw[decorate, red, thick] (0,-1)--(0,-2) node[right] {$J_2$};
\draw[decorate,red,thick] (0,1) -- (0,2) node[right] {$J_2$};
\end{tikzpicture}}
\qquad\qquad
\scalebox{0.8}{
\begin{tikzpicture}[baseline]
\tikzstyle{every node}=[font=\footnotesize]
\node[draw, circle] (node1) at (-1.5,0) {$N_{2}$};
\node[draw, circle] (node2) at (1.5,0) {$N_{2}$};
\node[draw, rectangle] (sqnode1) at (-3.5,0) {$n_1$};
\node[draw, rectangle] (sqnode2) at (3.5,0) {$n_2$}; 
\draw[draw=black,solid,thick,-]  (node1)--(sqnode1) ; 
\draw[draw=black,solid,thick,-]  (node2)--(sqnode2) ; 
\draw[draw=red,solid,thick,-,snake it]  (node1) to[bend left=20]  node[midway,above] {{\red $T(U(N))$}}  (node2) ; 
\draw[draw=red,solid,thick,-,snake it]  (node1) to[bend right=20]  node[midway,below] {{\red $T(U(N))$}}  (node2) ;
\end{tikzpicture}}
\ee
and focus on the cases of $N=1$ and $N=2$.  Such theories have interesting physical properties as we shall describe below.

Let us first discuss the abelian case. Since this theory admits a conventional Lagrangian description, we can easily analyse this theory along the line of \cite{Garozzo:2018kra}.  The detailed analysis is provided in Appendix \ref{Sec:ParaModuli}.  We find that whenever fundamental hypermultiplets are added to the quiver associated with parabolic $J$-folds, an interesting branch of the moduli space arises, mainly due to the presence of the gauge neutral monopole (or dressed monopole) operators.  In particular, for quiver \eref{flvpara} with $N=1$, we find that there are two branches of the moduli space. One can be identified as the Higgs branch and the other can be identified as the Coulomb branch, both of which are hyperK\"ahler cones. This feature is very similar to that of general 3d $\CN=4$ gauge theories.  The Higgs branch is isomorphic to a product of the closures of the minimal nilpotent orbits $\bar{\CO}^{SU(n_1)}_{\text{min}} \times \bar{\CO}^{SU(n_2)}_{\text{min}}$, where each factor is generated by the mesons constructed using the chiral multiplets in each fundamental hypermultiplet; see \eref{Higgsabel}.  The Coulomb branch is isomorphic to $\BC^2/\BZ_{n_1+n_2}$, which is generated by the monopole operators $V_{\pm(1,1)}$ with fluxes $\pm(1,1)$ and the complex scalar in the vector multiplet; see \eref{Coulabel}.  For a general $n_1$ and $n_2$, this theory has a global symmetry $\left( \frac{U(n_1) \times U(n_2)}{U(1)} \right) \times U(1)$, where the former factor denotes the flavour symmetry coming from the fundamental hypermultiplets and latter $U(1)$ denotes the topological symmetry.  For the special case of $n_1+n_2=2$, the $U(1)$ topological symmetry gets enhanced to $SU(2)$, which is also an isometry of the Coulomb branch $\BC^2/\BZ_2$.  Interestingly, if we set one of $n_1$ or $n_2$ to zero, say 
\be \label{twoTlinkswnflv}
\scalebox{0.8}{
\begin{tikzpicture}[baseline]
\tikzstyle{every node}=[font=\footnotesize]
\node[draw, circle] (node1) at (-1.5,0) {$1_{2}$};
\node[draw, circle] (node2) at (1.5,0) {$1_{2}$};
\node[draw, rectangle] (sqnode2) at (3.5,0) {$n$}; 
\draw[draw=black,solid,thick,-]  (node2)--(sqnode2) ; 
\draw[draw=red,solid,thick,-,snake it]  (node1) to[bend left=20]  node[midway,above] {{\red $T(U(1))$}}  (node2) ; 
\draw[draw=red,solid,thick,-,snake it]  (node1) to[bend right=20]  node[midway,below] {{\red $T(U(1))$}}  (node2) ;
\end{tikzpicture}}
\ee
This theory turns out to be the same as quiver \eref{TUNloop} with $N=1$, $k=2$, which is identical to 3d $\CN=4$ $U(1)$ gauge theory with $n$ flavours (\ie~ the $T_{(n,n-1)}(SU(n))$ theory \cite{Gaiotto:2008ak}).  One can indeed check that the moduli spaces and the indices of the two theories are equal.  Such an identification indicates that when $n_1=0$ (or $n_2=0$), the two $J_2$ duality walls can be ``collapsed'' into one, and the gauge node in \eref{twoTlinkswnflv} that is not flavoured can be removed such that the $T$-link becomes a loop around the other gauge node.   We remark that this statement only holds for the abelian case; we will see that for $N=2$ this is no longer true.

Quiver \eref{flvpara} with $N>1$ still bares the same features as in the abelian ($N=1$) theory.  In general, the index of \eref{flvpara} contains the terms $x^{\frac{1}{2}(n_1+n_2)}(w_1 w_2 + w_1^{-1} w_2^{-1})$, which indicates that there are gauge invariants monopole operators $V_{\pm (1,0,\ldots, 0; 1,0,\ldots,0)}$, with fluxes $\pm(1,0,\ldots,0)$ under each of the $U(N)$ gauge group, carrying $R$-charge $\frac{1}{2}(n_1+n_2)$.  Again, for $n_1+n_2=2$, the $U(1)$ topological symmetry gets enhanced to $SU(2)$.  Furthermore, when $n_1+n_2=1$, \ie~ $(n_1,n_2)=(1,0)$ or $(0,1)$, such monopole operators decouple as a free hypermultiplet (this is similar to the one flavour case discussed in section \ref{sec:oneflvU2k}). Let us consider, in particular, the case of $N=2$, $n_1=0$ and $n_2=1$:
\be \label{twoTlinksN2w1flv}
\scalebox{0.8}{
\begin{tikzpicture}[baseline]
\tikzstyle{every node}=[font=\footnotesize]
\node[draw, circle] (node1) at (-1.5,0) {$2_{2}$};
\node[draw, circle] (node2) at (1.5,0) {$2_{2}$};
\node[draw, rectangle] (sqnode2) at (3.5,0) {$1$}; 
\draw[draw=black,solid,thick,-]  (node2)--(sqnode2) ; 
\draw[draw=red,solid,thick,-,snake it]  (node1) to[bend left=20]  node[midway,above] {{\red $T(U(2))$}}  (node2) ; 
\draw[draw=red,solid,thick,-,snake it]  (node1) to[bend right=20]  node[midway,below] {{\red $T(U(2))$}}  (node2) ;
\end{tikzpicture}}
\ee
Indeed the index can be written as
\be
\CI_{\eref{twoTlinksN2w1flv}} = \CI_{\text{free}} (x; w) \times \CI^{\eref{twoTlinksN2w1flv}}_{\text{SCFT}} (x; w)
\ee
where we define $w$ as the product of the topological fugacities associated with the two gauge groups: $w= w_1 w_2$. The index of the free hypermultiplet $\CI_{\text{free}} (x; w)$ is given by \eref{freeindex}, and the index for the interacting SCFT is
\be \label{twoTlinkn1SCFT}
\begin{split}
\CI^{\eref{twoTlinksN2w1flv}}_{\text{SCFT}} (x; w) &= 1+x \chi^{SU(2)}_{[2]}(w) +x^2 \left[ \chi^{SU(2)}_{[4]}(w) - \chi^{SU(2)}_{[2]}(w) \right] \\
& \qquad -x^{\frac{5}{2}} \chi^{SU(2)}_{[2]}(w) +\ldots~.
\end{split}
\ee
with the unrefinement
\be
\CI^{\eref{twoTlinksN2w1flv}}_{\text{SCFT}} (x; w=1) = 1 + 3 x + 2 x^2 - 2 x^{5/2} - 4 x^3 + \ldots
\ee
The interacting SCFT has a flavour symmetry $SU(2)$.  Notice that the index of the SCFT \eref{twoTlinkn1SCFT} is different from \eref{indexintSCFTn1}.  (Hence, we cannot collapse two $J_2$ duality walls into one as in the abelian case.) In particular, while \eref{indexintSCFTn1} exhibits the presence of the extra-SUSY current multiplet, \eref{twoTlinkn1SCFT} does not.

\section{Gauge group $SU(2)_k/\BZ_2$} \label{sec:SU2modZ2}
In this section we consider a theory with a single $SU(2)_k/\BZ_2$ gauge node with $n$ $T$-links attached to it.  
\be \label{TUNloopSU2kmodZ2}
\scalebox{0.8}{
\begin{tikzpicture}[baseline]
\tikzstyle{every node}=[font=\footnotesize]
\node[draw, circle] (node1) at (0,0) {$SU(2)_k/\BZ_2$};
\draw[red,thick] (node1) edge [out=135,in=225,loop,looseness=3, snake it]  (node1);
\draw[red,thick] (node1) edge [out=45,in=135,loop,looseness=3, snake it]  (node1);
\draw[red,thick] (node1) edge [out=45,in=-45,loop,looseness=3, snake it]  (node1);
\node[draw=none] at (0,-1.3) {\red \huge $\ldots$}; 
\node[draw=none] at (0,-1.8) {\red $n$ T-links} ;
\end{tikzpicture}}
\ee
Although the brane configuration for $n >1$ is not known, we demonstrate below that such theories have interesting properties from the field theoretic perspective.

The indices involving $SU(N)/\BZ_N$ gauge group for theories in 3d were discussed in \cite{Fazzi:2018rkr, Razamat:2014pta}\footnote{Note the index for 3d gauge theories can be obtained as the limit of the lens space index for 4d gauge theories \cite{Benini:2011nc}.  As discussed extensively in \cite{Razamat:2013opa}, the latter is sensitive to the global structure of the gauge group.}.  Here, we write down the expression of the index for \eref{TUNloopSU2kmodZ2}, analogous to those presented in \cite{Fazzi:2018rkr}:\footnote{Since $SU(2)/\BZ_2$ is isomorphic to $SO(3)$, one can also compute the index for $SO(3)$ gauge group using the formulae described in \cite[sec. 6.1]{Aharony:2013kma}.  Note that the normalisation for the CS level for $SO(3)$ is such that $SO(3)_k = SU(2)_{2k}/\BZ_2$, and that the fugacity $\zeta$ for the $\BZ_2^{\mathcal{M}}$ symmetry in \cite[sec. 6.1]{Aharony:2013kma} is identified with the fugacity $g$ here.}
\be
\begin{split}
 \CI_{\eref{TUNloopSU2kmodZ2}}(x; g) &= \frac{1}{2} \sum_{l=0}^1 g^l \sum_{m \in \BZ + \frac{l}{2}} \oint \frac{d z}{2 \pi i z} x^{-2|m|} \prod_{\pm} \left( 1- (-1)^{2m} z^{\pm 2} x^{2|m|} \right)  \\
& \qquad \times z^{2k m} \left[ \hat{\CI}_{T(SU(2))}(x; \{z, m\}, \{z, m \})\right]^n~,
\end{split}
\ee
where $g$ is a fugacity for the global $\BZ_2$ symmetry which takes values $1$ or $-1$, and $\hat{\CI}_{\text{$U(1)$ with 2 flv}}$ is the index for the $U(1)$ gauge theory with $2$ flavours such that the sum over the gauge flux is properly quantised:
\be
\begin{split}
&\hat{\CI}_{T(SU(2))}(x; \{\mu, p\}, \{\tau, n\}) \\
&= \sum_{m\in \BZ+\frac{1}{2}(p \, \mod \, 2)} \tau^{2m} \,  \oint\frac{\text{d}z}{2\pi i z}z^{2n} x^{\frac{|m-p|}{2}}\frac{((-1)^{m-p}z^{\mp1}\mu^{\pm1}x^{3/2+|m-p|;x^2})_\infty}{((-1)^{m-p}z^{\pm1}\mu^{\mp1}x^{1/2+|m-p|;x^2})_\infty}\times\\
&\qquad x^{\frac{|m+p|}{2}}\frac{((-1)^{m+p}z^{\mp1}\mu^{\mp1}x^{3/2+|m+p|;x^2})_\infty}{((-1)^{m+p}z^{\pm1}\mu^{\pm1}x^{1/2+|m+p|;x^2})_\infty}~.
\end{split}
\ee  

We may obtain the result for the gauge group $SU(2)_{k}$, instead of $SU(2)_{k}/\BZ_2$, by gauging the $\BZ_2$ global symmetry associated with $g$:
\be \label{gaugingZ2}
 \CI_{\text{\eref{TUNloopSU2kmodZ2} with $SU(2)_k$ gauge group}} = \frac{1}{2} \left[\CI_{\eref{TUNloopSU2kmodZ2}}(x; g=1)+ \CI_{\eref{TUNloopSU2kmodZ2}}(x; g=-1) \right]~.
\ee

We find that for $k \in \BZ$ and $n=1$, the index is the same as that of the theory with the same $k$ presented in section \ref{sec:U2kwnoflv}.  However, the result becomes more interesting when both $n$ and $k$ is even, since the index depends on the fugacity $g$. This indicates the presence of the operators carrying a non-trivial charge under the $\BZ_2$ discrete symmetry which are gauge invariant monopole operators.  Let us focus on the case of $n=2$.  We provide some examples in the following table.
\begin{longtable}{|c|c|}
\hline
CS level & Index  \\
\hline
$k=0$ & $1 + 2 g x + 4 x^2 - 4 x^3 + (9 + 8 g) x^4+\ldots$ \\
\rowcolor{yellow} $|k|=2$ & $1 + (2 - g) x^2 - (4 + 4 g) x^3 + (9 + 8 g) x^4+\ldots$ \\
$|k|=4$ & $1 + g x + 3 x^2 - (4 + 2 g) x^3 + (8 + 8 g) x^4+\ldots$\\
$|k|=6$ & $1 + (2 + g) x^2 - (4 + 4 g) x^3 + (7 + 4 g) x^4+\ldots$ \\
$|k|=8$ & $1 + 2 x^2 - 4 x^3 + (7 - 4 g) x^4+\ldots$ \\
\hline
\end{longtable}
\noindent The case of $|k|=2$ is highlighted in yellow to indicate that the index exhibits supersymmetry enhancement.  Since the coefficient of $x$ is zero, the theory has no flavour current.  The term $-g$ at order $x^2$ indicates the presence of an extra SUSY-current multiplet, acted non-trivially by the $\BZ_2$ global symmetry. For this reason, we conclude that supersymmetry is enhanced to $\CN=4$.  We emphasise that it is important to refine the index with respect to $g$ in order to see such a multiplet. On the other hand, the indices for the other values of $k$ do not exhibit the existence of the extra SUSY-current multiplet.  The same is also true if we gauge the $\BZ_2$ global symmetry as described in \eref{gaugingZ2}. 

Finally, let us consider the case in which $k$ is half-odd-integral and $n=1$.  For $|k| \geq \frac{1}{2}$, we find that the indices are different from those theories that have been considered in earlier, and so it seems to us that these theories are new.  Moreover, they exhibit the presence of an extra SUSY-current multiplet, which leads to the conclusion that supersymmetry is enhanced to $\CN=4$.  We tabulate the indices for a few values of half-odd-integral CS levels below.
\begin{longtable}{|c|l|}
\hline
CS level & Index  \\
\hline
$|k|=1/2$ & $1 - x^2 + 2 x^3 - x^4 - 2 x^5+6 x^6+\ldots$ \\
 $|k|=3/2$ & $1 - x^2 + 2 x^3 - 2 x^5 + 6 x^6+\ldots$ \\
$|k|=5/2$ & $1 - x^2 + 2 x^3 - 2 x^4 - 2 x^5 + 6 x^6+\ldots$ \\
\hline
\end{longtable}

\section{Conclusion and perspectives} \label{sec:conclusion}
Several properties of 3d $S$-fold SCFTs have been investigated using supersymmetric indices.  We have found several theories whose indices exhibit supersymmetry enhancement, due to the presence of extra-supersymmetry current multiplets. Dualities between different $S$-fold quiver theories have also been explored and the indices allow us to establish the operator map between such theories.  Moreover, we studied $S$-fold theories whose gauge symmetries have different global structures, namely $SU(2)$ and $SU(2)/\BZ_2$.  We found that the indices of the latter reveal interesting properties regarding the discrete topological symmetry as well as supersymmetry enhancement in a certain case.

These findings lead to a number of future directions of research.  First of all, there are a number of theories in this paper whose indices do not exhibit the presence of extra-SUSY current multiplets.  It would be interesting to study the amount of supersymmetry that is possessed by these theories in more detail.  It would also be nice to study the marginal $\CN=2$ deformations of the $S$-fold SCFTs along the line of the recent paper \cite{Bachas:2019jaa}.  Another interesting point is to understand better new interacting SCFTs whose indices are computing in \eref{indexintSCFTn1} and \eref{twoTlinkn1SCFT}.  Finally, it would be interesting to explore 3d SCFTs associated with duality walls in other theories than 4d $\CN=4$ super-Yang-Mills and $\CN=2^*$ theories.  Explicit Lagrangian descriptions for a number of such 3d theories have been proposed, for example, in \cite{Floch:2015hwo, Benini:2017dud} for the $S$-duality wall of the 4d $\CN=2$ $SU(N)$ gauge theory with $2N$ flavours.  There are also another class of theories that have been recently studied in \cite{Pasquetti:2019uop, Pasquetti:2019tix}.  It would be interesting to study whether such theories still have similar properties to quiver theories that contain $T$-link associated with $T(U(N))$.

\acknowledgments
We would like to thank Antonio Amariti, Constantin Bachas, Luca Martucci, Sara Pasquetti, Giulia Peveri, and especially Chiung Hwang, Alessandro Tomasiello and Alberto Zaffaroni for very helpful comments and a number of useful discussions. M.\,S.\,  is partially supported by the ERC-STG grant 637844-HBQFTNCER and by the INFN.


\appendix
\section{Moduli space of flavoured abelian parabolic $J$-fold theories}
\label{Sec:ParaModuli}
In this section we study the moduli space of a class of parabolic $J$-fold theories, in the presence of the hypermultiplet fundamental matter.  We focus on the models with abelian gauge group, since the Lagrangian description is available.  More general detailed discussions can be found in \cite{Garozzo:2018kra}.

For definiteness, let us focus on the following model with $U(1)_{k_1} \times U(1)_{k_2}$ gauge group:
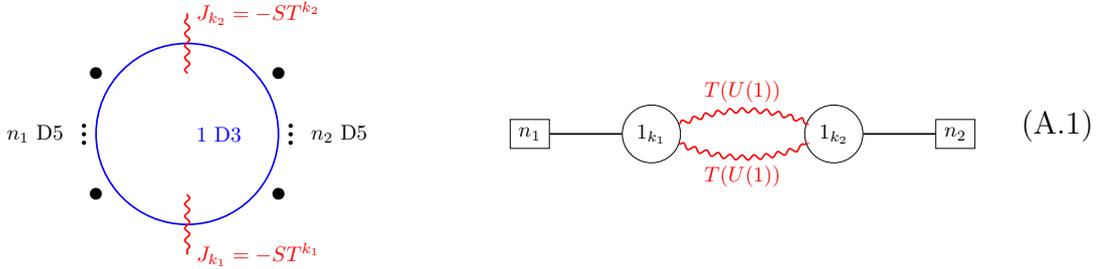
\be
\label{2U(1)2T}
\scalebox{0.8}{
\begin{tikzpicture}[baseline]
\tikzstyle{every node}=[font=\footnotesize]
\draw[blue,thick] (0,0) circle (1.5cm) node[midway, right] {$1$ D3};
\node[draw=none] at (-1.5,1) {\large{$\bullet$}};
\node[draw=none] at (-1.7,0.1) {\Huge{$\vdots$}};
\node[draw=none] at (-2.5,0) {$n_1$ D5};
\node[draw=none] at (-1.5,-1) {\large{$\bullet$}};
\node[draw=none] at (1.5,1) {\large{$\bullet$}};
\node[draw=none] at (1.7,0.1) {\Huge{$\vdots$}};
\node[draw=none] at (2.5,0) {$n_2$ D5};
\node[draw=none] at (1.5,-1) {\large{$\bullet$}};
\tikzset{decoration={snake,amplitude=.4mm,segment length=2mm,
                       post length=0mm,pre length=0mm}}
\draw[decorate, red, thick] (0,-1)--(0,-2) node[right] {$J_{k_1}=-ST^{k_1}$};
\draw[decorate,red,thick] (0,1) -- (0,2) node[right] {$J_{k_2}=-ST^{k_2}$};
\end{tikzpicture}}
\qquad\qquad
\scalebox{0.8}{
\begin{tikzpicture}[baseline]
\tikzstyle{every node}=[font=\footnotesize]
\node[draw, circle] (node1) at (-1.5,0) {$1_{k_1}$};
\node[draw, circle] (node2) at (1.5,0) {$1_{k_2}$};
\node[draw, rectangle] (sqnode1) at (-3.5,0) {$n_1$};
\node[draw, rectangle] (sqnode2) at (3.5,0) {$n_2$}; 
\draw[draw=black,solid,thick,-]  (node1)--(sqnode1) ; 
\draw[draw=black,solid,thick,-]  (node2)--(sqnode2) ; 
\draw[draw=red,solid,thick,-,snake it]  (node1) to[bend left=20]  node[midway,above] {{\red $T(U(1))$}}  (node2) ; 
\draw[draw=red,solid,thick,-,snake it]  (node1) to[bend right=20]  node[midway,below] {{\red $T(U(1))$}}  (node2) ;
\end{tikzpicture}}
\ee
For the moment we allow for generic CS levels $k_1$ and $k_2$, but we will see that the vacuum equations admit solutions for non-trivial branches of the moduli space when $J_1 J_2$ is parabolic, \ie ~$|\tr J_1 J_2|=2$, or equivalently $k_1k_2\,=0$ or $4$. 

Let us rewrite the quiver \eqref{2U(1)2T} in $\cN=2$ language:
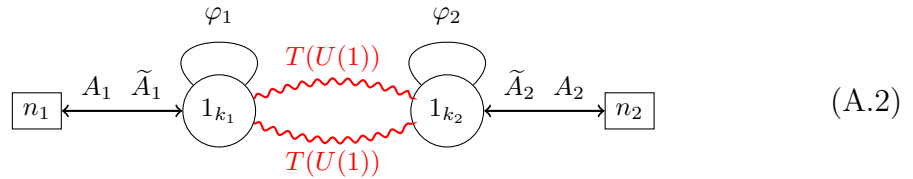
\be
\scalebox{1}{
\begin{tikzpicture}[baseline]
\tikzstyle{every node}=[font=\footnotesize]
\node[draw, circle] (node1) at (-1.5,0) {$1_{k_1}$};
\node[draw, circle] (node2) at (1.5,0) {$1_{k_2}$};
\node[draw, rectangle] (sqnode1) at (-3.9,0) {$n_1$};
\node[draw, rectangle] (sqnode2) at (3.9,0) {$n_2$}; 
\draw[draw=black,solid,thick,<->]  (node1) to node[midway,above] {$A_1\,\,\,\,\tilde A_1$} (sqnode1); 
\draw[draw=black,solid,thick,<->]  (node2) to node[midway,above] {$\tilde A_2\,\,\,\,A_2$} (sqnode2); 
\draw[black] (node1) edge [out=45,in=135,loop,looseness=4] node[midway,above=0.1cm] {$\varphi_1$}  (node1);
\draw[black] (node2) edge [out=45,in=135,loop,looseness=4] node[midway,above=0.1cm] {$\varphi_2$}  (node2);
\draw[draw=red,solid,thick,-,snake it]  (node1) to[bend left=20]  node[midway,above] {{\red $T(U(1))$}}  (node2) ; 
\draw[draw=red,solid,thick,-,snake it]  (node1) to[bend right=20]  node[midway,below] {{\red $T(U(1))$}}  (node2) ;
\end{tikzpicture}}
\ee
with superpotential:
\be
W\,=\, -\tr(A_1\varphi_1 \tilde A_1+A_2\varphi_2 \tilde A_2)\,+\, \frac{1}{2}(k_1 \varphi_1^2+k_2\varphi_2^2)\,{\color{blue}-2\varphi_1\varphi_2}\,.
\ee
where we denoted in blue the contribution due to the two $T$-links, consisting of a mixed CS coupling. The vacuum equations are as follows:
\be
\label{F2}
A_1\varphi_1\,=\,\tilde A_1\varphi_1\,=\,0\,,\quad A_2\varphi_2\,=\,\tilde A_2\varphi_2\,=\,0\,,
\ee
and
\be
\begin{split} 
\label{F1}
&k_1 \varphi_1\,{\color{blue}-2\varphi_2}\,=\, (A_1)_a (\tilde A_1)^a \,,\\
&k_2 \varphi_2\,{\color{blue} -2\varphi_1}\,=\, (A_2)_i (\tilde A_2)^i.\\
\end{split}
\ee
where $a,b,c=1,\ldots, n_1$ and $i,j,k =1, \ldots, n_2$.

The vacuum equations \eqref{F2} and \eqref{F1} admit the solutions in which $\varphi_1=\varphi_2=0$, regardless of the CS levels.  This branch of the moduli space is generated by the mesons $(M_1)^b_a = (A_{1})_a (\tilde{A}_1)^b$ and $(M_2)^j_i = (A_{2})_i (\tilde{A}_2)^j$ subject to the following relations:
\be
\mathrm{rank}(M_{1,2}) \leq 1~, \qquad M^2_{1,2}=0~,
\ee
where the first relations come from the fact that each of the matrices $M_{1}$ and $M_2$ is constructed as a product of two vectors, and the second matrix relations follow from \eref{F1}. We refer to this branch of the moduli space as the Higgs branch, denoted by $\cH_{\eqref{2U(1)2T}}$.  Indeed, it is isomorphic to a product of the closures of the minimal nilpotent orbits:
\be \label{Higgsabel}
\cH_{\eqref{2U(1)2T}}\,=\bar{\CO}^{SU(n_1)}_{\min} \times \bar{\CO}^{SU(n_2)}_{\min}.
\ee
There are also other non-trivial branches of moduli spaces, which we are analysing in the following.  

Let us consider the branch on which $\varphi_1 \neq 0$ and $\varphi_2 \neq 0$.  From \eref{F2}, we have $A_1=\tilde A_1=A_2=\tilde A_2=0$. Equations \eqref{F1} admit solutions only if:
\be \label{k1k24}
k_1\varphi_1\,=2\varphi_2\,, \qquad k_2 \varphi_2 = 2\varphi_1\,, \qquad k_1k_2-4\,=\,0\,;
\ee
the latter implies that $J_1 J_2$ has to be parabolic such that either $(k_1,k_2)=(1,4)$ or $(k_1,k_2)=(2,2)$. (The case the $(k_1,k_2)=(4,1)$ can be considered by simply exchanging $n_1$ and $n_2$.)  We analyse these cases below.
\bi
\item {\bf The case of $(k_1, k_2)=(2,2)$.} The first equation of \eref{k1k24} sets $\varphi_1=\varphi_2\equiv \varphi$. Since the real scalars in the vector multiplets belong to the same multiplets as $\varphi_{1,2}$, the magnetic fluxes of the monopole operators $V_{(m_1,m_2)}$ satisfy $m_1=m_2 \equiv m$.  The $R$-charge and the gauge charges with respect to the first and second nodes are respectively
\be
\begin{split}
&R[V_{(m,m)}]\,=\,\frac{1}{2}(n_1+n_2)|m|\,,\\
&q_1[V_{(m,m)}]\,=-(k_1m - 2m) =0\,,\quad q_2[V_{(m,m)}]= -(k_2 m -2m) =0\,.
\end{split}
\ee
Observe that the $V_{(m,m)}$ are gauge neutral for all $m$. This branch is generated by the basic monopole operators $V_{\pm(1,1)}$ and $\varphi$ (the latter has $R$-charge $1$), satisfying the quantum relation.
\be
V_{(1,1)}\,V_{-(1,1)}\,=\,\varphi^{n_1+n_2}~.
\ee
This branch is thus a Coulomb branch and it is isomorphic to 
\be \label{Coulabel}
\mathcal{C}^{k_1=k_2=2}_{\eqref{2U(1)2T}}\,=\,\mathbb{C}^2/\mathbb{Z}_{n_1+n_2}\,.
\ee
In the special case of one flavour, \ie~ $(n_1,n_2)=(1,0)$ or $(0,1)$, we see that the Coulomb branch is isomorphic to $\BC^2 \cong \BH$.  Indeed, the basic monopole operators decouple as a free hypermultiplet.
\item {\bf The case of $(k_1, k_2)=(1,4)$.} In this case $\varphi_1 = 2 \varphi_2 = 2 \varphi$ and the allowed magnetic fluxes for the monopole operators $V_{(m_1,m_2)}$ are such that $m_1=2m_2 \equiv 2m$.  The $R$-charge and the gauge charges with respect to the first and second nodes are respectively
\be
\begin{split}
&R[V_{(2m,m)}]\,=\,\frac{1}{2}(n_1 |2m| + n_2 |m|) = \left(n_1+\frac{1}{2} n_2 \right)|m|\\
&q_1[V_{(2m,m)}]\,=-[k_1(2m) - 2m] =0\,,\quad q_2[V_{(2m,m)}]= -[k_2 (m) -2(2m)] =0\,.
\end{split}
\ee
Observe that $V_{(2m,m)}$ are gauge neutral for all $m$.  This branch of the moduli space is generated by $V_{\pm (2,1)}$ and $\varphi$, satisfying the quantum relation:
\be
V_{(2,1)} V_{-(2,1)} = \varphi^{2n_1+ n_2}~.
\ee
This branch is thus a Coulomb branch and it is isomorphic to 
\be
\mathcal{C}^{(k_1,k_2)=(1,4)}_{\eqref{2U(1)2T}}\,=\,\mathbb{C}^2/\mathbb{Z}_{2n_1+n_2}\,.
\ee 
\ei
It is worth pointing out that for both $(k_1, k_2) =(2,2)$ and $(1,4)$, the vacuum equations admit the solutions such that there is a clear separation between the Higgs and Coulomb branches, in the same way as general 3d $\CN=4$ gauge theories.  This is mainly due to the fact that the monopole operators are gauge neutral.  Note also that both branches are hyperK\"ahler cones.

Next, we analyse the case in which one of $\varphi_1$ and $\varphi_2$ is zero.  For definiteness, let us take $\varphi_2=0$ and $0 \neq \varphi_1 \equiv \varphi$.  From \eref{F2}, we see that $A_1 = \tilde{A}_1=0$, and so \eref{F1} admits a solution only if $k_1=0$.  Let us suppose that 
\be
k_1=0~.
\ee  
Observe that the CS levels $(0,k_2)$ satisfies the parabolic condition on $J_{0} J_{k_2} $, because $|\Tr(J_0 J_{k_2})| =2$ for any $k_2$.
Then the second equation of \eref{F1} implies that 
\be
(A_2)_i (\tilde{A}_2)^i= -2\varphi ~.
\ee
The fluxes $(m_1,m_2)$ of the monopole operators $V_{(m_1,m_2)}$, satisfies $m_2=0$.  For convenience, we write $m_1=m$.  The $R$-charge and the gauge charges of the monopole operators $V_{(m,0)}$ are 
\be
\begin{split}
&R[V_{(m,0)}]\,=\,\frac{1}{2}(n_1 |m| + n_2 |0|) = \frac{1}{2} n_1|m|\\
&q_1[V_{(m,0)}]\,=-[k_1(m) - 2(0)] = 0 \,,\quad q_2[V_{(m,0)}]= -[k_2 (0) -2(m)] =2m\,.
\end{split}
\ee
In this case the monopole operator $V_{(m,0)}$ is no longer neutral under the gauge symmetry, but it carries charge $2m$ under the $U(1)_{k_2}$ gauge group.  We can form the basic gauge invariant dressed monopole operators as follows:
\be \label{dressedmon0kabel}
(W^+)^{ij} = V_{(1,0)} (\tilde{A}_2)^i  (\tilde{A}_2)^j~, \qquad (W^-)_{ij} = V_{(-1,0)} (A_2)_i (A_2)_j~.
\ee
These operators transform under the representation $[2,0, \ldots, 0]$ and $[0, \ldots, 0,2]$ of $SU(n_2)$ respectively. The carries $R$-charges
\be
R[W^\pm] = \frac{1}{2}n_1 +1~,
\ee
and satisfy the quantum relation
\be
\Tr (W^+ W^-) = (W^+)^{ij} (W^-)_{ji} = \varphi^{n_1+2}~.
\ee
Since the dressed monopole operators $W^\pm$ are generators of this branch of the moduli space, we can regard this as a ``mixed'' Higgs and Coulomb branch.   

Note that if we take instead $\varphi_1=0$ and $0 \neq \varphi_2 \equiv \varphi$, the situation is reversed. In order for the vacuum equations to admit a solution we must have $k_2=0$.  This leads to the gauge invariant dressed monopole operators
\be
(U^+)^{ij} = V_{(0,1)} (\tilde{A}_1)^i  (\tilde{A}_1)^j~, \qquad (U^-)_{ij} = V_{(0,-1)} (A_1)_i (A_1)_j~,
\ee
which transform under the representation $[2,0, \ldots, 0]$ and $[0, \ldots, 0,2]$ of $SU(n_1)$ respectively.  The carries $R$-charges $R[U^\pm] = \frac{1}{2}n_2 +1$ and satisfy the quantum relation $\Tr (U^+ U^-) = \varphi^{n_2+2}$.

Finally, we remark that if $(k_1,k_2)=(0,0)$, which is another possibility for $J_{k_1} J_{k_2}$ to be parabolic, then both dressed monopole operators $W^\pm$ and $U^\pm$, as described above, are generators of the moduli space.


\bibliographystyle{ytphys}
\bibliography{ref}
\end{document}